\documentclass[12pt]{article}
\usepackage{amsmath,amssymb}
\usepackage{bm}
\usepackage{mathrsfs}
\usepackage{fourier}
\usepackage{multirow}
\allowdisplaybreaks[4]
\newcommand{{\SlashD}}{D\!\!\!\!\!\!\big/}
\newcommand{{\Slashq}}{q\!\!\!\!\!\big/}
\newcommand{{\SlashF}}{{\rm F}\!\!\!\!\!/}
\usepackage[usenames]{color}

\begin{document}

\title{A bottom-up approach to fermion mass hierarchy:
a case with vector-like fermions}

\author{
Yoshiharu \textsc{Kawamura}\footnote{E-mail: haru@azusa.shinshu-u.ac.jp}\\
{\it Department of Physics, Shinshu University, }\\
{\it Matsumoto 390-8621, Japan}\\
}

\date{
September 9, 2019}

\maketitle
\begin{abstract}
We propose a bottom-up approach that a structure of a high-energy physics
is explored by accumulating existence proofs and/or no-go theorems 
in the standard model or its extension.
As an illustration, we study fermion mass hierarchies 
based on an extension of the standard model
with vector-like fermions.
It is shown that a magnitude of elements of Yukawa coupling matrices can become $O(1)$
and a Yukawa coupling unification can be realized in a theory beyond the extended model, 
if vector-like fermions mix with three families.
In this case, small Yukawa couplings in the standard model can be highly sensitive
to a small variation of matrix elements,
and it seems that the mass hierarchy occurs
as a result of a fine tuning.
\end{abstract}

\section{Introduction}

One of the most fascinating riddles in particle physics
is the origin of the fermion mass hierarchy and flavor mixing
in the standard model (SM).
Various intriguing attempts have been performed to solve it.
Most of them are based on the top-down approach~\cite{CEG,F,HHW,F2,GJ,FN}.
Starting from high-energy theories (HETs) 
such as grand unified theories and superstring theories
or extensions of the SM with some flavor symmetries,
Yukawa coupling matrices are constructed or ansatzes
called texture zeros are proposed, 
to explain the flavor structure in the SM.
In spite of endless efforts, we have not arrived at satisfactory answers,
because a theory beyond the SM has not yet been confirmed and
there is no powerful guiding principle to determine it.
Flavor symmetries are possible candidates, and
the flavor structure of quarks and leptons has been studied
intensively, based on various flavor symmetries~\cite{FN,MY,I,HPS,HS,AF,IKOSOT,IKOOST}.
It is shown that there are no exact flavor-dependent symmetries in the SM~\cite{LNS,Koide}.

In the exploration of the flavor physics, 
the bottom-up approach has also been carried out~\cite{Ku,AN,St,DR,YK,YK2}.
Observed values of fermion masses and mixing angles
become a springboard for study on the origin of flavor structure.
In some cases, the analyses are made based on specific models.
In other cases, the form of Yukawa coupling matrices can be specified
to some extent by adopting a guiding principle and/or taking reasonable assumptions,
independent of models in HETs,
in the framework of the SM or its extension.
This approach has also a limitation,
because global U(3) symmetries exist in the fermion kinetic terms,
Yukawa coupling matrices contain unphysical parameters,
and the structure of HETs cannot be
completely identified from experimental data alone.
However, it can offer useful information on HETs, depending on how it is used.
In concrete,  
we make plausible conjectures 
in the extension with extra particles and/or under additional assumptions,
and then we can give some statements
as theorems, by examining whether they are correct or not.

In this paper, we propose a bottom-up approach that a structure of a high-energy physics
is explored by accumulating existence proofs and/or no-go theorems 
in the standard model or its extension.
Based on the bottom-up approach, 
we make conjectures on Yukawa couplings 
and pursue whether they are realized or not
in an extension of the SM including heavy vector-like fermions,
without specifying HETs beyond the extension.
We consider two conjectures.
One is that {\it a magnitude of Yukawa couplings can become $O(1)$ in a HET}.
It comes from our expectation
that the magnitude of dimensionless parameters on terms 
allowed by symmetries should be $O(1)$ in a fundamental theory.
If it is true,
the hierarchical structure of Yukawa couplings occurs 
after a transition from a HET to the extension of the SM
or through a mechanism in some lower-energy physics.
In Appendix A, we explain the backgrounds of this conjecture.
The other is that {\it Yukawa couplings can be unified in a HET}.
It stems from a symmetry principle. 
It is deeply related to 
a grand unification based on SO(10)~\cite{SO10} and E$_6$~\cite{E6}.
These conjectures provide boundary conditions at the HET scale.
One of our goals is to present
our strategy, its availability, and its limitation
of our approach,
and hence we focus on the quark sector, in the following.

The outline of this paper is as follows.
In the next section, we introduce our strategy based on the SM
and explain our setup on an extension of the SM.
In Sect. 3, we examine whether above-mentioned conjectures hold or not
in the extended one.
In the last section, we give conclusions and discussions.

\section{An extension of the standard model}

\subsection{Strategy}

We take a bottom-up approach that an origin of flavor structure in the SM
is explored by compiling no-go theorems.
Before we explain our setup on an extension of the SM, 
we introduce our strategy using the SM.
Let us start with the quark sector in the SM, described by
the Lagrangian density:
\begin{eqnarray}
{\mathscr{L}}_{\rm SM}^{\rm quark} = \overline{q}_{{\rm L}i} i \SlashD q_{{\rm L}i}
+ \overline{u}_{{\rm R}i} i \SlashD u_{{\rm R}i}
+ \overline{d}_{{\rm R}i} i \SlashD d_{{\rm R}i}
 - y_{ij}^{(u)} \overline{q}_{{\rm L}i} \tilde{\phi} u_{{\rm R}j}
- y_{ij}^{(d)} \overline{q}_{{\rm L}i} \phi d_{{\rm R}j} + {\rm h.c.},
\label{L-ky-quark}
\end{eqnarray}
where $q_{{\rm L}i}$ are left-handed quark doublets,
$u_{{\rm R}i}$ and $d_{{\rm R}i}$ are
right-handed up- and down-type quark singlets,
$i, j (=1, 2, 3)$ are family labels, 
summation over repeated indices is understood with few exceptions
throughout this paper,
$y_{ij}^{(u)}$ and $y_{ij}^{(d)}$ are Yukawa coupling matrices,
$\phi$ is the Higgs doublet, $\tilde{\phi} = i \tau_2 \phi^*$
and h.c. stands for hermitian conjugation of former terms.

Here, we assume that there exists a set of privileged field variables
($q'_{\rm L}$, $u'_{\rm R}$, $d'_{\rm R}$)
relating to the flavor structure.
A candidate is a unitary basis of flavor symmetries~\cite{YK,YK2}.
By the change of field variables as
\begin{eqnarray}
&~& q_{\rm L} = N_q  q'_{\rm L},~~u_{\rm R} = N_u u'_{\rm R},~~d_{\rm R} = N_d d'_{\rm R},
\label{unitary-SM}
\end{eqnarray}
the above Lagrangian density is rewritten by
\begin{eqnarray}
&~& {\mathscr{L}'}_{\rm SM}^{\rm quark} 
= k_{ij}^{(q)} \overline{q}'_{{\rm L}i} i \SlashD q'_{{\rm L}j}
+ k_{ij}^{(u)} \overline{u}'_{{\rm R}i} i \SlashD u'_{{\rm R}j}
+ k_{ij}^{(d)} \overline{d}'_{{\rm R}i} i \SlashD d'_{{\rm R}j}
\nonumber \\
&~& ~~~~~~~~~~~~~~~~~~~~ 
- \left(y_1\right)_{ij} \overline{q}'_{{\rm L}i} \tilde{\phi} u'_{{\rm R}j}
- \left(y_2\right)_{ij} \overline{q}'_{{\rm L}i} \phi d'_{{\rm R}j} + {\rm h.c.},
\label{L-SM-quark-prime}
\end{eqnarray}
where $N_q$, $N_u$, and $N_d$ are $3 \times 3$ complex matrices,
$k_{ij}^{(q)}$, $k_{ij}^{(u)}$, and $k_{ij}^{(d)}$ are quark kinetic coefficients,
and $\left(y_1\right)_{ij}$ and $\left(y_2\right)_{ij}$ are Yukawa coupling
matrices for privileged fields.
They yield the relations:
\begin{eqnarray}
\hspace{-0.7cm}
&~& k^{(q)}_{ij} = \left(N_{q}^{\dagger}N_{q}\right)_{ij},~~
k^{(u)}_{ij} = \left(N_{u}^{\dagger}N_{u}\right)_{ij},~~
k^{(d)}_{ij} = \left(N_{d}^{\dagger}N_{d}\right)_{ij},~~
\label{k-I}\\
\hspace{-0.7cm}
&~& \left(y_1\right)_{ij} = \left(N_{q}^{\dagger} y^{(u)} N_{u}\right)_{ij}
= \left(N_{q}^{\dagger} {V_{\rm L}^{(u)}}^{\dagger} 
y_{\rm diag}^{(u)} V_{\rm R}^{(u)} N_{u}\right)_{ij},~~
\label{y1SM}\\
\hspace{-0.7cm}
&~& \left(y_2\right)_{ij} = \left(N_{q}^{\dagger} y^{(d)} N_{d}\right)_{ij}
= \left(N_{q}^{\dagger} {V_{\rm L}^{(d)}}^{\dagger} 
y_{\rm diag}^{(d)} V_{\rm R}^{(d)} N_{d}\right)_{ij}
= \left(N_{q}^{\dagger} {V_{\rm L}^{(u)}}^{\dagger} V_{\rm KM} 
y_{\rm diag}^{(d)} V_{\rm R}^{(d)} N_{d}\right)_{ij},
\label{y2SM}
\end{eqnarray}
where $V_{\rm L}^{(u)}$, $V_{\rm L}^{(d)}$, $V_{\rm R}^{(u)}$, and $V_{\rm R}^{(d)}$
are unitary matrices and, using them, the Yukawa coupling matrices are diagonalized as
\begin{eqnarray}
V_{\rm L}^{(u)} y^{(u)} {V_{\rm R}^{(u)}}^{\dagger}
= y_{\rm diag}^{(u)} = {\rm diag}\left(y_u, y_c, y_t\right),~~
V_{\rm L}^{(d)} y^{(d)} {V_{\rm R}^{(d)}}^{\dagger}
= y_{\rm diag}^{(d)} = {\rm diag}\left(y_d, y_s, y_b\right).
\label{y-diag}
\end{eqnarray}
$V_{\rm KM}$ is the Kobayashi-Maskawa matrix defined by~\cite{KM}
\begin{eqnarray}
V_{\rm KM} \equiv V_{\rm L}^{(u)} {V_{\rm L}^{(d)}}^{\dagger}.
\label{VKM}
\end{eqnarray}
Using experimental values of quark masses,
$y_{\rm diag}^{(u)}$, $y_{\rm diag}^{(d)}$,
and $V_{\rm KM}$ are roughly estimated 
at the weak scale as\cite{PDG}
\begin{eqnarray}
&~& y_{\rm diag}^{(u)} = 
{\rm diag}\left(1.3 \times 10^{-5},~ 7.3 \times 10^{-3},~ 1.0\right) 
= {\rm diag}\left(\lambda^7, \lambda^4, 1\right),
\label{yu-diag-value}\\
&~& y_{\rm diag}^{(d)} 
= {\rm diag}\left(2.7 \times 10^{-5},~ 
5.5 \times 10^{-4},~ 2.4 \times 10^{-2}\right)
= {\rm diag}\left(\lambda^7, \lambda^5, \lambda^3\right),
\label{yd-diag-value}\\
&~&  V_{\rm KM} = \left(
\begin{array}{ccc}
1 & \lambda & \lambda^3 \\
\lambda & 1 & \lambda^2 \\
\lambda^3 & \lambda^2 & 1
\end{array}
\right).
\label{VKM-lambda}
\end{eqnarray}
In the final expressions, $\lambda^n$ means 
$\displaystyle{O\left(\lambda^n\right)}$
with $\lambda = \sin\theta_{\rm C} \cong 0.225$ 
($\theta_{\rm C}$ is the Cabibbo angle~\cite{C}).\footnote{
Although the magnitude of $\displaystyle{\left(V_{\rm KM}\right)_{13}}$
is $0.00365 \pm 0.00012$ and is regarded as $\displaystyle{O\left(\lambda^4\right)}$,
we treat it as $\displaystyle{O\left(\lambda^3\right)}$
with respect for the Wolfenstein parametrization~\cite{Wolf}.
}
Physical parameters, in general, receive radiative corrections,
and the above values should be evaluated by considering 
renormalization effects to match with their counterparts of HET at a high-energy scale.

${\mathscr{L}'}_{\rm SM}^{\rm quark}$
is supposed to be obtained as a result of a transition from a theory
beyond the SM and can possess useful information
on what is behind the flavor structure. 
Note that non-canonical matter kinetic terms, in general,
appear in ${\mathscr{L}'}_{\rm SM}^{\rm quark}$,
as will be explained in Appendix A.\footnote{
Several works on the flavor physics have been carried out
based on non-canonical matter kinetic terms~\cite{LNS2,BD1,BD2,BLR,KY,HKY,KP,EI,KPRVV,Liu,DIU}.
}

Because the kinetic coefficients are hermitian and positive definite, 
they are written by
\begin{eqnarray}
k_{ij}^{(q)} = \left(U_q^{\dagger} \left(J_{q}\right)^2 U_q\right)_{ij},~~
k_{ij}^{(u)} = \left(U_u^{\dagger} \left(J_{u}\right)^2 U_u\right)_{ij},~~
k_{ij}^{(d)} = \left(U_d^{\dagger} \left(J_{d}\right)^2 U_d\right)_{ij},
\label{k-UJ}
\end{eqnarray}
where $U_q$, $U_u$, and $U_d$ are $3 \times 3$ unitary matrices, 
and $J_q$, $J_u$, and $J_d$ are real $3 \times 3$ diagonal matrices.
Then, $N_q$, $N_u$, and $N_d$ are expressed by
\begin{eqnarray}
N_q = V_q J_q U_q,~~ N_u = V_u J_u U_u,~~ N_d = V_d J_d U_d,
\label{NqNuNd}
\end{eqnarray}
where $V_q$, $V_u$, and $V_d$ are $3 \times 3$ unitary matrices.
As a magnitude of elements in unitary matrices is not beyond $1$,
we obtain the inequalities:
\begin{eqnarray}
\left(J_q\right)_{ii} \le O(1),~~ \left(J_u\right)_{ii} \le O(1),~~ 
\left(J_d\right)_{ii} \le O(1),~~ ({\rm no~~\!summation~~\!on}~~i)
\label{JqJuJd}
\end{eqnarray}
from the requirement that the magnitude of matter kinetic coefficients
should be at most $O(1)$ and the appearance of suppression factors 
in terms containing higher-dimensional operators.
Then, from Eqs. (\ref{y2SM}) and (\ref{yd-diag-value}), 
we have the following no-go theorem.\\
$[$Theorem$]$ {\it If $k_{ij}^{(q)} \le O(1)$ and $k_{ij}^{(d)} \le O(1)$,
the magnitude of elements of down-type Yukawa coupling matrices cannot go beyond
$\displaystyle{O\left(10^{-2}\right)}$, i.e., 
$\displaystyle{\left(y_2\right)_{ij} \le O\left(\lambda^3\right)}$, 
in the framework of SM}.

For reference, in a case with a large mixing such as 
$\displaystyle{\left(N_q\right)_{ij} = O(1)}$ 
and $\displaystyle{\left(N_d\right)_{ij} = O(1)}$,
the magnitude of several components
in $\displaystyle{\left(y_2\right)_{ij}}$ can be
$\displaystyle{O\left(\lambda^3\right)}$. 
In contrast, the magnitude of the top-quark Yukawa coupling is sizable as $y_t = 1.0$,
and that of various components
in $\displaystyle{\left(y_1\right)_{ij}}$ can be $O(1)$
through a large mixing such as $\displaystyle{\left(N_q\right)_{ij} = O(1)}$ 
and $\displaystyle{\left(N_u\right)_{ij} = O(1)}$.
Here and hereafter, we regard $O(1)$ as a number 
greater than $\lambda(\cong 0.225)$ and less than $\lambda^{-1}(\cong 4.44)$, i.e.,
$O(1)$ contains $1/\sqrt{2}$ and $1/2$.

Next, we examine whether the Yukawa couplings can be unified or not.
Using ${\mathscr{L}'}_{\rm SM}^{\rm quark}$, 
we obtain the following no-go theorem.\\
$[$Theorem$]$ {\it Under the kinetic unification such as 
$k_{ij}^{(q)} = k_{ij}^{(u)} = k_{ij}^{(d)}$,
the exact Yukawa coupling unification such as
$\displaystyle{\left(y_1\right)_{ij} = \left(y_2\right)_{ij}}$ does not occur
in the framework of SM}.

It is understood from the observation that 
we obtain unrealistic features 
such that Yukawa coupling matrices have same eigenvalues,
degenerate masses are derived between up- and down-type quarks, 
and the Kobayashi-Maskawa matrix becomes a unit matrix,
if the kinetic coefficients are common and
the up-type Yukawa coupling matrix is identical to the down-type one.

In the case that the unifications are required from a symmetry, 
symmetry breaking terms appear and ruin some unification relations  
in the broken phase.
Here, we consider the case that the kinetic unification is destroyed
with $k_{ij}^{(u)} \ne k_{ij}^{(d)}$.
In this case, from Eqs. (\ref{y1SM}) and (\ref{y2SM}), 
the following relation should hold
\begin{eqnarray}
\left(N^{\dagger}_q  {V_{\rm L}^{(u)}}^{\dagger}
y_{\rm diag}^{(u)} V_{\rm R}^{(u)} N_{u}\right)_{ij}
= \left(N^{\dagger}_q  {V_{\rm L}^{(u)}}^{\dagger} V_{\rm KM} 
y_{\rm diag}^{(d)} V_{\rm R}^{(d)} N_{d}\right)_{ij},
\label{y1=y2SM}
\end{eqnarray}
to realize $\displaystyle{\left(y_1\right)_{ij} = \left(y_2\right)_{ij}}$.
Using Eqs. (\ref{yu-diag-value}), (\ref{yd-diag-value}), (\ref{VKM-lambda}), and
(\ref{y1=y2SM}),
we obtain the relation:
\begin{eqnarray}
V_{\rm R}^{(u)} N_{u}
= \left(y_{\rm diag}^{(u)}\right)^{-1} V_{\rm KM} y_{\rm diag}^{(d)}
V_{\rm R}^{(d)} N_{d}
= \left(
\begin{array}{ccc}
\lambda^0 & \lambda^{-1} & \lambda^{-1} \\
\lambda^4 & \lambda & \lambda \\
\lambda^{10} & \lambda^{7} & \lambda^3
\end{array}
\right)V_{\rm R}^{(d)} N_{d}.
\label{Nu-Nd}
\end{eqnarray}
When the magnitude of $\displaystyle{\left(N_{u}\right)_{ij}}$ is $O(1)$,
we have $\displaystyle{\left(V_{\rm R}^{(d)} N_{d}\right)_{2j} \le O(\lambda)}$ 
and $\displaystyle{\left(V_{\rm R}^{(d)} N_{d}\right)_{3j} \le O(\lambda)}$,
and then every element becomes small, i.e., 
$\displaystyle{\left(y_1\right)_{ij} = \left(y_2\right)_{ij} 
\le O\left(\lambda^4\right)}$.
Then, the magnitude of 
every element in $k_{ij}^{(u)}$ and $k_{ij}^{(d)}$ becomes $O(1)$.
Hence the Yukawa unification can occur, and
it would be also interesting to study a Yukawa unification model such that
a magnitude of all elements of 
original Yukawa coupling matrices can be much smaller than $O(1)$.

In this way, we have obtained no-go theorems on the Yukawa couplings
using experimental data
in the framework of SM, without specifying HETs.
Through such a down-to-earth approach, 
knowledge and information on the flavor structure are expected to be accumulated,
and some clues to the origin of flavor and hints on HETs are provided.

In the following, we apply this strategy in an extension of the SM.

\subsection{Setup}

We adopt several assumptions.\\
(a) A theory beyond the SM, which is referred to as HET,
has higher gauge symmetries.
It owns a seed of the flavor structure, and
the form of Yukawa coupling matrices is determined by HET.
Fermion kinetic terms do not necessarily take a canonical form,
where the origin of flavor structure is unveiled~\cite{YK,YK2}.\\
(b) At a high-energy scale, the theory turns out to be an extension of the SM, i.e.,
a model with the SM gauge group 
G$_{\rm SM}(=$ SU(3)$_{\rm C} \times$ SU(2)$_{\rm L} \times$ U(1)$_{\rm Y}$)
and extra particles.
We refer to it as ``SM + $\alpha$''.
One should be careful not to confuse SM + $\alpha$ with HET.\footnote{
In our terminology,
grand unified theories belong to HET, and the minimal 
supersymmetric extension of the SM (MSSM)
belongs to SM + $\alpha$.
}\\
(c) The extra particles have large masses, compared with the weak boson mass, and
the SM particles survive after the decoupling of heavy ones.
There are 4th generation fermions and their mirror particles, as extra particles.
Here, mirror particles are particles with opposite quantum numbers
under G$_{\rm SM}$. 
A fermion and its mirror one obey a vector representation in pairs,
and hence they are often referred to as a vector-like fermion.

We consider the Lagrangian density of quarks in SM + $\alpha$,
described by
\begin{eqnarray}
&~& {\mathscr{L}'}_{\rm SM+\alpha}^{\rm quark} 
= k_{IJ}^{(q)} \overline{q'_{{\rm L}I}} i \SlashD q'_{{\rm L}J}
+ k_{IJ}^{(u)} \overline{u'_{{\rm R}I}} i \SlashD u'_{{\rm R}J}
+ k_{IJ}^{(d)} \overline{d'_{{\rm R}I}} i \SlashD d'_{{\rm R}J}
\nonumber \\
&~&  ~~~~~~~~~~~~~~~~~~ 
+ k^{(q_{\rm m})} \overline{\left(q'_{{\rm L(m)}}\right)^{c}} 
i \SlashD \left(q'_{{\rm L(m)}}\right)^c
+ k^{(u_{\rm m})} \overline{\left(u'_{{\rm R(m)}}\right)^c}
i \SlashD \left(u'_{{\rm R(m)}}\right)^c
\nonumber \\
&~&  ~~~~~~~~~~~~~~~~~~ 
+ k^{(d_{\rm m})} \overline{\left(d'_{{\rm R(m)}}\right)^c} 
i \SlashD \left(d'_{{\rm R(m)}}\right)^c
\nonumber \\
&~&  ~~~~~~~~~~~~~~~~~~ 
- y^{(U)}_{IJ} \overline{q'_{{\rm L}I}} \tilde{\phi} u'_{{\rm R}J}
- y^{(D)}_{IJ} \overline{q'_{{\rm L}I}} \phi d'_{{\rm R}J} + {\rm h.c.}
\nonumber \\
&~&  ~~~~~~~~~~~~~~~~~~ 
- y^{(u_{\rm m})} \overline{\left(u'_{{\rm R(m)}}\right)^{c}} 
\left(q'_{{\rm L(m)}}\right)^c \tilde{\phi}^* 
- y^{(d_{\rm m})} \overline{\left(d'_{{\rm R(m)}}\right)^{c}} 
\left(q'_{{\rm L(m)}}\right)^c {\phi}^* + {\rm h.c.}  
\nonumber \\
&~&  ~~~~~~~~~~~~~~~~~~ 
- m_I^{(q_{\rm m})} \overline{q'_{{\rm L}I}} \left(q'_{{\rm L(m)}}\right)^c
- m_J^{(u_{\rm m})} \overline{\left(u'_{{\rm R(m)}}\right)^{c}} u'_{{\rm R}J}
- m_J^{(d_{\rm m})} \overline{\left(d'_{{\rm R(m)}}\right)^{c}} d'_{{\rm R}J}
 + {\rm h.c.},
\label{L-SM+alpha-quark}
\end{eqnarray}
where $q'_{{\rm L}I}$ are counterparts of left-handed quark doublets,
$u'_{{\rm R}I}$ and $d'_{{\rm R}I}$ are those of
right-handed up- and down-type quark singlets, and
$I$ and $J$ run from 1 to 4.
$\displaystyle{\left(q'_{{\rm L(m)}}\right)^c}$,
$\displaystyle{\left(u'_{{\rm R(m)}}\right)^c}$,
and $\displaystyle{\left(d'_{{\rm R(m)}}\right)^c}$
are charge conjugations of mirror quarks $q'_{{\rm L(m)}}$, $u'_{{\rm R(m)}}$,
and $d'_{{\rm R(m)}}$.
We refer to $q'_{{\rm L}I}$, $u'_{{\rm R}I}$, $d'_{{\rm R}I}$,
$\displaystyle{\left(q'_{{\rm L(m)}}\right)^c}$,
$\displaystyle{\left(u'_{{\rm R(m)}}\right)^c}$,
and $\displaystyle{\left(d'_{{\rm R(m)}}\right)^c}$
as quarks, collectively.
Fields with prime represents privileged fields concerning the flavor structure.
$k_{IJ}^{(q)}$, $k_{IJ}^{(u)}$, $k_{IJ}^{(d)}$,
$k^{(q_{\rm m})}$, $k^{(u_{\rm m})}$, and $k^{(d_{\rm m})}$
are quark kinetic coefficients,
$y^{(U)}_{IJ}$, $y^{(D)}_{IJ}$, $y^{(u_{\rm m})}$, and $y^{(d_{\rm m})}$
are Yukawa coupling matrices, and
$m_I^{(q_{\rm m})}$, $m_J^{(u_{\rm m})}$, and $m_J^{(d_{\rm m})}$
are mass parameters.

The gauge quantum numbers and the chirality $\gamma_5$ of quarks 
are listed in Table \ref{T-SM+alpha}.
For reference, we list the gauge quantum numbers 
and the chirality of mirror quarks 
in Table \ref{T-mirrors}.
In Tables \ref{T-SM+alpha} and \ref{T-mirrors}, Y is the weak hypercharge.
\begin{table}[htbp]
\caption{The gauge quantum numbers and the chirality
of quarks in SM + $\alpha$.}
\label{T-SM+alpha}
\begin{center}
\begin{tabular}{c|c|c|c|c}
\hline                              
quarks in SM + $\alpha$ & SU(3)$_{\rm C}$ & SU(2)$_{\rm L}$ & Y & $\gamma_5$
\\ \hline 
$q'_{{\rm L}I} = \left(
\begin{array}{c}
u'_{{\rm L}I} \\
d'_{{\rm L}I}
\end{array} 
\right)$ & $\bm{3}$ & $\bm{2}$ & $\frac{1}{6}$ & $-1$\\
$\left(q'_{{\rm L(m)}}\right)^c= \left(
\begin{array}{c}
\left(u'_{{\rm L(m)}}\right)^c \\
\left(d'_{{\rm L(m)}}\right)^c
\end{array} 
\right)$ & $\bm{3}$ & $\bm{2}$ & $\frac{1}{6}$ & $1$\\
$u'_{{\rm R}I}$ & $\bm{3}$ & $\bm{1}$ & $\frac{2}{3}$ & $1$\\
$\left(u'_{{\rm R(m)}}\right)^c$ & $\bm{3}$ & $\bm{1}$ & $\frac{2}{3}$ & $-1$\\
$d'_{{\rm R}I}$ & $\bm{3}$ & $\bm{1}$ & $-\frac{1}{3}$ & $1$\\
$\left(d'_{{\rm R(m)}}\right)^c$ & $\bm{3}$ & $\bm{1}$ & $-\frac{1}{3}$ & $-1$\\
\hline
\end{tabular}
\end{center}
\end{table}

\begin{table}[htbp]
\caption{The gauge quantum numbers and the chirality
of mirror quarks in SM + $\alpha$.}
\label{T-mirrors}
\begin{center}
\begin{tabular}{c|c|c|c|c}
\hline                              
mirror quarks & SU(3)$_{\rm C}$ & SU(2)$_{\rm L}$ & Y & $\gamma_5$
\\ \hline 
$q'_{{\rm L(m)}}= \left(
\begin{array}{c}
u'_{{\rm L(m)}} \\
d'_{{\rm L(m)}}
\end{array} 
\right)$ & $\overline{\bm{3}}$ & $\bm{2}$ & $-\frac{1}{6}$ & $-1$\\
$u'_{{\rm R(m)}}$ & $\overline{\bm{3}}$ & $\bm{1}$ & $-\frac{2}{3}$ & $1$\\
$d'_{{\rm R(m)}}$ & $\overline{\bm{3}}$ & $\bm{1}$ & $\frac{1}{3}$ & $1$\\
\hline
\end{tabular}
\end{center}
\end{table}

The Yukawa interactions and mass terms in ${\mathscr{L}'}_{\rm SM+\alpha}^{\rm quark}$
are compactly written by
\begin{eqnarray}
{\mathscr{L}'}_{\rm SM+\alpha (Y, m)}^{\rm quark}
= - \overline{U'_{{\rm L}A}} M^{(U)}_{AB} U'_{{\rm R}B}
 - \overline{D'_{{\rm L}A}} M^{(D)}_{AB} D'_{{\rm R}B}
 + {\rm h.c.} + \cdots,
\label{L-Y+M}
\end{eqnarray}
where $A$ and $B$ run from 1 to 5, and
$U'_{{\rm L}A}$, $U'_{{\rm R}A}$, $D'_{{\rm L}A}$,
and $D'_{{\rm R}A}$ consist of 5 components such that
\begin{eqnarray}
U'_{{\rm L}} =  
\left(
\begin{array}{c}
u'_{{\rm L}I} \\
\left(u'_{{\rm R(m)}}\right)^c
\end{array} 
\right),~~
U'_{{\rm R}} =  
\left(
\begin{array}{c}
u'_{{\rm R}I} \\
\left(u'_{{\rm L(m)}}\right)^c
\end{array} 
\right),~~
D'_{{\rm L}} =  
\left(
\begin{array}{c}
d'_{{\rm L}I} \\
\left(d'_{{\rm R(m)}}\right)^c
\end{array} 
\right),~~
D'_{{\rm R}} =  
\left(
\begin{array}{c}
d'_{{\rm R}I} \\
\left(d'_{{\rm L(m)}}\right)^c
\end{array} 
\right),
\label{U+D}
\end{eqnarray}
and $M^{(U)}_{AB}$ and $M^{(D)}_{AB}$ are $5 \times 5$ complex matrices given by
\begin{eqnarray}
M^{(U)} = 
\left(
\begin{array}{cc}
y_{IJ}^{(U)} \phi^{0*} & m_I^{(q_{\rm m})} \\
m_J^{(u_{\rm m})} & y^{(u_{\rm m})} \phi^{0}
\end{array} 
\right),~~
M^{(D)} = 
\left(
\begin{array}{cc}
y_{IJ}^{(D)} \phi^{0} & m_I^{(q_{\rm m})} \\
m_J^{(d_{\rm m})} & y^{(d_{\rm m})} \phi^{0*}
\end{array} 
\right),
\label{M(U)}
\end{eqnarray}
respectively.
The ellipsis in Eq. (\ref{L-Y+M}) stands for terms 
containing a charged component of the Higgs doublet.

To apply the bottom-up approach to our model,
we need to know the relationship between
the privileged fields in ${\mathscr{L}'}_{\rm SM+\alpha}^{\rm quark}$ 
and mass eigenstates in the SM.
In the following, we will see that 
the kinetic terms change into the canonical form
and mass terms including Yukawa interactions are diagonalized
by redefining field variables.

First, we pay attention to the fact that the quark kinetic coefficients
are hermitian and are expressed by
\begin{eqnarray}
&~& K^{(U_{\rm L})} = 
\left(
\begin{array}{cc}
k_{IJ}^{(q)} & 0 \\
0 & k^{(u_{\rm m})}
\end{array} 
\right) = {N_{\rm L}^{u}}^{\dagger}N_{\rm L}^{u},~~
K^{(D_{\rm L})} = 
\left(
\begin{array}{cc}
k_{IJ}^{(q)} & 0 \\
0 & k^{(d_{\rm m})}
\end{array} 
\right) = {N_{\rm L}^{d}}^{\dagger}N_{\rm L}^{d},~~
\nonumber \\
&~& K^{(U_{\rm R})} = 
\left(
\begin{array}{cc}
k_{IJ}^{(u)} & 0 \\
0 & k^{(q_{\rm m})}
\end{array} 
\right) = {N_{\rm R}^{u}}^{\dagger}N_{\rm R}^{u},~~
K^{(D_{\rm R})} = 
\left(
\begin{array}{cc}
k_{IJ}^{(d)} & 0 \\
0 & k^{(q_{\rm m})}
\end{array} 
\right) = {N_{\rm R}^{d}}^{\dagger}N_{\rm R}^{d},
\label{K(UL)}
\end{eqnarray}
using $5 \times 5$ complex matrices
$N_{\rm L}^{u}$, $N_{\rm L}^{d}$, $N_{\rm R}^{u}$, and $N_{\rm R}^{d}$
with $\displaystyle{\left(N_{\rm L}^{u}\right)_{IJ}= \left(N_{\rm L}^{d}\right)_{IJ}}$,
$\displaystyle{\left(N_{\rm R}^{u}\right)_{55}= \left(N_{\rm R}^{d}\right)_{55}}$,
$\displaystyle{\left(N_{\rm L}^{u}\right)_{I5}= 0}$,
$\displaystyle{\left(N_{\rm L}^{u}\right)_{5J}= 0}$ and so on.
Then, after the change of variables:
\begin{eqnarray}
&~& U_{\rm L} =  
\left(
\begin{array}{c}
u_{{\rm L}I} \\
\left(u_{{\rm R(m)}}\right)^c
\end{array} 
\right) = N_{\rm L}^{u} U'_{\rm L},~~
D_{\rm L} =  
\left(
\begin{array}{c}
d_{{\rm L}I} \\
\left(d_{{\rm R(m)}}\right)^c
\end{array} 
\right) = N_{\rm L}^{d} D'_{\rm L},~~
\label{UL}\\
&~& U_{\rm R}=  
\left(
\begin{array}{c}
u_{{\rm R}I} \\
\left(u_{{\rm L(m)}}\right)^c
\end{array} 
\right) = N_{\rm R}^{u} U'_{\rm R},~~
D_{\rm R} =  
\left(
\begin{array}{c}
d_{{\rm R}I} \\
\left(d_{{\rm L(m)}}\right)^c
\end{array} 
\right) = N_{\rm R}^{d} D'_{\rm R},
\label{DL}
\end{eqnarray}
we obtain the canonical type of quark kinetic terms:
\begin{eqnarray}
&~& {\mathscr{L}}_{\rm SM+\alpha (k)}^{\rm quark} 
= \overline{q_{{\rm L}I}} i \SlashD q_{{\rm L}I}
+ \overline{u_{{\rm R}I}} i \SlashD u_{{\rm R}I}
+ \overline{d_{{\rm R}I}} i \SlashD d_{{\rm R}I}
\nonumber \\
&~&  ~~~~~~~~~~~~~~~~~~~~~ 
+ \overline{\left(q_{{\rm L(m)}}\right)^{c}} 
i \SlashD \left(q_{{\rm L(m)}}\right)^c
+ \overline{\left(u_{{\rm R(m)}}\right)^c}
i \SlashD \left(u_{{\rm R(m)}}\right)^c
+ \overline{\left(d_{{\rm R(m)}}\right)^c} 
i \SlashD \left(d_{{\rm R(m)}}\right)^c,
\label{L-kinetic}
\end{eqnarray}
where $q_{{\rm L}I}$ and $\displaystyle{\left(q_{{\rm L(m)}}\right)^c}$ 
are SU(2)$_{\rm L}$ doublets given by
\begin{eqnarray}
q_{{\rm L}I} =  
\left(
\begin{array}{c}
u_{{\rm L}I} \\
d_{{\rm L}I}
\end{array} 
\right),~~
\left(q_{{\rm L(m)}}\right)^c =  
\left(
\begin{array}{c}
\left(u_{{\rm L(m)}}\right)^c \\
\left(d_{{\rm L(m)}}\right)^c
\end{array} 
\right),
\label{qqc}
\end{eqnarray}
respectively.

Here, we give comments.
The transformation matrices
$N_{\rm L}^{u}$, $N_{\rm L}^{d}$, $N_{\rm R}^{u}$, and $N_{\rm R}^{d}$
are not completely fixed, or
$\tilde{V}_{\rm L}^{u} N_{\rm L}^{u}$, $\tilde{V}_{\rm L}^{d}N_{\rm L}^{d}$, 
$\tilde{V}_{\rm R}^{u} N_{\rm R}^{u}$, and $\tilde{V}_{\rm R}^{d} N_{\rm R}^{d}$
also offer the same kinetic coefficients
$K^{(U_{\rm L})}$, $K^{(D_{\rm L})}$, $K^{(U_{\rm R})}$, 
and $K^{(D_{\rm R})}$, respectively.
Here, $\tilde{V}_{\rm L}^{u}$, 
$\tilde{V}_{\rm L}^{d}$, $\tilde{V}_{\rm R}^{u}$, and $\tilde{V}_{\rm R}^{d}$
are arbitrary unitary matrices, and using this arbitrariness,
$M^{(U)}$ and $M^{(D)}$ are diagonalized, as will be described below.
Then, $V_{\rm KM}$
appears in $\displaystyle{\overline{q_{{\rm L}I}} i \SlashD q_{{\rm L}I}}$
on the mass eigenstates.
The quark kinetic terms in ${\mathscr{L}'}_{\rm SM+\alpha}^{\rm quark}$
cannot be compactly written
by using $U'_{{\rm L}}$, $U'_{{\rm R}}$, $D'_{{\rm L}}$,
and $D'_{{\rm R}}$, because these variables contain fields with 
different quantum numbers under SU(2)$_{\rm L} \times$ U(1)$_{\rm Y}$
and $u'_{{\rm L}I}$, $d'_{{\rm L}I}$,
$\displaystyle{\left(u'_{{\rm L(m)}}\right)^c}$,
and $\displaystyle{\left(d'_{{\rm L(m)}}\right)^c}$
are not treated as SU(2)$_{\rm L}$ doublets in them.

Next, by the change of variables such as Eqs. (\ref{UL}) and (\ref{DL}),
including suitable unitary matrices
$\tilde{V}_{\rm L}^{u}$, $\tilde{V}_{\rm L}^{d}$, 
$\tilde{V}_{\rm R}^{u}$, and $\tilde{V}_{\rm R}^{d}$,
$M^{(U)}$ and $M^{(D)}$ are transformed into 
\begin{eqnarray}
&~& \left({N_{\rm L}^{u}}^{\dagger}\right)^{-1} M^{(U)} 
\left({{N}}_{\rm R}^{u}\right)^{-1}
= \left(
\begin{array}{ccccc}
y_{u}{\phi}^{0*} & 0 & 0 & 0 & 0 \\
0 & y_{c}{\phi}^{0*} & 0 & 0 & 0 \\
0 & 0 & y_{t}{\phi}^{0*} & 0 & 0 \\
0 & 0 & 0 & 0 & m_1^{(U)} \\
0 & 0 & 0 & m_2^{(U)} & 0
\end{array} 
\right),
\label{MU-diag} \\
&~& \left({N_{\rm L}^{d}}^{\dagger}\right)^{-1} M^{(D)} 
\left({{N}}_{\rm R}^{d}\right)^{-1}
= \left(
\begin{array}{ccccc}
y_{d}\phi^{0} & 0 & 0 & 0 & 0 \\
0 & y_{s}\phi^{0} & 0 & 0 & 0 \\
0 & 0 & y_{b}\phi^{0} & 0 & 0 \\
0 & 0 & 0 & 0 & m_1^{(D)} \\
0 & 0 & 0 & m_2^{(D)} & 0
\end{array} 
\right),
\label{MD-diag}
\end{eqnarray}
where, for simplicity,
$\tilde{V}_{\rm L}^{u} N_{\rm L}^{u}$, $\tilde{V}_{\rm L}^{d}N_{\rm L}^{d}$, 
$\tilde{V}_{\rm R}^{u} N_{\rm R}^{u}$, and $\tilde{V}_{\rm R}^{d} N_{\rm R}^{d}$
are again denoted as $N_{\rm L}^{u}$, $N_{\rm L}^{d}$, 
$N_{\rm R}^{u}$, and $N_{\rm R}^{d}$, respectively.
$m_1^{(U)}$, $m_2^{(U)}$, $m_1^{(D)}$, and $m_2^{(D)}$ 
are large masses of extra quarks.
The experimental bounds on
4th generation quark masses
require that an extra up-type quark is heavier than 1160~GeV from neutral-current decays
and an extra down-type quark is heavier than 880~GeV from charged-current decays~\cite{PDG}.

\section{Examination on conjectures}

We carry out order estimations using $\lambda (\cong 0.225)$,
and examine whether the conjectures on the Yukawa couplings hold or not,
based on the extension of the SM described by 
${\mathscr{L}'}_{\rm SM+\alpha}^{\rm quark}$.
The analyses on the case with partial multiplets are
carried out in Appendix B.

\subsection{Seeking transformation matrices}

Using $\lambda$, 
the magnitude of the right-hand sides 
in Eqs. (\ref{MU-diag}) and (\ref{MD-diag}) 
is parametrized as
\begin{eqnarray}
M^{(U)}_{\rm diag}
= \left(
\begin{array}{ccccc}
\lambda^7 & 0 & 0 & 0 & 0 \\
0 & \lambda^4 & 0 & 0 & 0 \\
0 & 0 & \lambda^0 & 0 & 0 \\
0 & 0 & 0 & 0 & \lambda^{-n_1} \\
0 & 0 & 0 & \lambda^{-n_2} & 0
\end{array} 
\right) \frac{v}{\sqrt{2}},~~
M^{(D)}_{\rm diag}
= \left(
\begin{array}{ccccc}
\lambda^7 & 0 & 0 & 0 & 0 \\
0 & \lambda^5 & 0 & 0 & 0 \\
0 & 0 & \lambda^3 & 0 & 0 \\
0 & 0 & 0 & 0 & \lambda^{-n_3} \\
0 & 0 & 0 & \lambda^{-n_4} & 0
\end{array} 
\right) \frac{v}{\sqrt{2}},
\label{M-diag-rep}
\end{eqnarray}
where $n_1$, $n_2$, $n_3$, and $n_4$ are positive integers,
as seen from the lower mass bounds 1160~GeV and 880~GeV,
and $v/\sqrt{2}$ is the vacuum expectation value of 
a neutral component of the Higgs doublet.
Note that $v$ is used for the sake of convenience,
although $m_1^{(U)}$, $m_2^{(U)}$, $m_1^{(D)}$, and $m_2^{(D)}$
must be irrelevant to the breakdown of electroweak symmetry.

Using Eqs. (\ref{M(U)}), (\ref{MU-diag}), (\ref{MD-diag}), and (\ref{M-diag-rep}), 
we obtain the relations:
\begin{eqnarray}
&~& 
\left(
\begin{array}{cc}
y_{IJ}^{(U)} \langle \phi^{0*} \rangle & m_I^{(q_{\rm m})} \\
m_J^{(u_{\rm m})} & y^{(u_{\rm m})} \langle \phi^{0} \rangle
\end{array} 
\right)
= {N_{\rm L}^{u}}^{\dagger} 
\left(
\begin{array}{ccccc}
\lambda^7 & 0 & 0 & 0 & 0 \\
0 & \lambda^4 & 0 & 0 & 0 \\
0 & 0 & \lambda^0 & 0 & 0 \\
0 & 0 & 0 & 0 & \lambda^{-n_1} \\
0 & 0 & 0 & \lambda^{-n_2} & 0
\end{array} 
\right) N_{\rm R}^{u} \frac{v}{\sqrt{2}},
\label{M(U)-diag}\\
&~& 
\left(
\begin{array}{cc}
y_{IJ}^{(D)} \langle \phi^{0} \rangle & m_I^{(q_{\rm m})} \\
m_J^{(d_{\rm m})} & y^{(d_{\rm m})} \langle \phi^{0*} \rangle
\end{array} 
\right)
= {N_{\rm L}^{d}}^{\dagger}
\left(
\begin{array}{ccccc}
\lambda^7 & 0 & 0 & 0 & 0 \\
0 & \lambda^5 & 0 & 0 & 0 \\
0 & 0 & \lambda^3 & 0 & 0 \\
0 & 0 & 0 & 0 & \lambda^{-n_3} \\
0 & 0 & 0 & \lambda^{-n_4} & 0
\end{array} 
\right) N_{\rm R}^{d} \frac{v}{\sqrt{2}}.
\label{M(D)-diag}
\end{eqnarray}

According to the backgrounds explained in Appendix A, 
we impose the following conditions on the kinetic coefficients,
\begin{eqnarray}
&~& k_{IJ}^{(q)},~~ k_{IJ}^{(u)},~~ k_{IJ}^{(d)} = O(1)~~~ {\rm for}~I=J,~~~
(I, J = 1, \cdots, 4)
\nonumber \\
&~& 
k_{IJ}^{(q)},~~ k_{IJ}^{(u)},~~ k_{IJ}^{(d)} \le O(1)~~~ {\rm for}~I \ne J,~~
\nonumber \\
&~& k^{(q_{\rm m})},~~ k^{(u_{\rm m})},~~ k^{(d_{\rm m})} = O(1).
\label{k=O(1)}
\end{eqnarray}
The conjectures on the Yukawa couplings are written by
\begin{eqnarray}
y^{(U)}_{IJ},~~ y^{(D)}_{IJ},~~ y^{(u_{\rm m})},~~ y^{(d_{\rm m})} = O(1)~~~
{\rm for~~\!some~~\!entries}
\label{y=O(1)}
\end{eqnarray}
and
\begin{eqnarray}
y^{(U)}_{IJ} = y^{(D)}_{IJ},
\label{yU=yD-IJ}
\end{eqnarray}
respectively.

The examination on conjectures is carried out 
by studying whether $N_{\rm L}^u$, $N_{\rm L}^d$, $N_{\rm R}^u$,
and $N_{\rm R}^d$ exist or not, to satisfy Eqs. (\ref{y=O(1)}) and (\ref{yU=yD-IJ}),
based on Eqs. (\ref{K(UL)}), (\ref{M(U)-diag}), (\ref{M(D)-diag}), and (\ref{k=O(1)}),
and specifying the form of those matrices.

Let us take the ansatzes:\footnote{
Although a generality is lost by adopting them, 
it is enough if $N_{\rm L}^u$, $N_{\rm L}^d$, $N_{\rm R}^u$,
and $N_{\rm R}^d$ are found with this choice,
from the viewpoint of a possible existence.
}
\begin{eqnarray}
&~& \left(N_{\rm L}^u\right)_{AB},~~ \left(N_{\rm L}^d\right)_{AB},~~ 
\left(N_{\rm R}^u\right)_{AB},~~ \left(N_{\rm R}^d\right)_{AB} = O(1)
~~~ {\rm for}~A=B,~~~ (A, B = 1, \cdots, 5)
\nonumber \\
&~& \left(N_{\rm L}^u\right)_{AB},~~ \left(N_{\rm L}^d\right)_{AB},~~ 
\left(N_{\rm R}^u\right)_{AB},~~ \left(N_{\rm R}^d\right)_{AB} \le O(1)
~~~ {\rm for}~A \ne B.
\label{N=O(1)}
\end{eqnarray}

First, the relations (\ref{K(UL)}) yield the conditions:
\begin{eqnarray}
&~& K_{I5}^{(U_{\rm L})} = \left({N_{\rm L}^u}^{\dagger}\right)_{IA}
\left(N_{\rm L}^u\right)_{A5} = 0,~~
K_{5J}^{(U_{\rm L})} = \left({N_{\rm L}^u}^{\dagger}\right)_{5A}
\left(N_{\rm L}^u\right)_{AJ} = 0,~~
\label{KI5=0-UL}\\
&~& K_{I5}^{(D_{\rm L})} = \left({N_{\rm L}^d}^{\dagger}\right)_{IA}
\left(N_{\rm L}^d\right)_{A5} = 0,~~
K_{5J}^{(D_{\rm L})} = \left({N_{\rm L}^d}^{\dagger}\right)_{5A}
\left(N_{\rm L}^d\right)_{AJ} = 0,~~
\label{KI5=0-DL}\\
&~& K_{I5}^{(U_{\rm R})} = \left({N_{\rm R}^u}^{\dagger}\right)_{IA}
\left(N_{\rm R}^u\right)_{A5} = 0,~~
K_{5J}^{(U_{\rm R})} = \left({N_{\rm R}^u}^{\dagger}\right)_{5A}
\left(N_{\rm R}^u\right)_{AJ} = 0,~~
\label{KI5=0-UR}\\
&~& K_{I5}^{(D_{\rm R})} = \left({N_{\rm R}^d}^{\dagger}\right)_{IA}
\left(N_{\rm L}^d\right)_{A5} = 0,~~
K_{5J}^{(D_{\rm R})} = \left({N_{\rm R}^d}^{\dagger}\right)_{5A}
\left(N_{\rm R}^d\right)_{AJ} = 0.
\label{KI5=0-DR}
\end{eqnarray}
These conditions are satisfied with suitable components,
if the rank of $4 \times 4$ sub-matrices $\left(N_{\rm L}^u\right)_{IJ}$,
$\left(N_{\rm L}^d\right)_{IJ}$, $\left(N_{\rm R}^u\right)_{IJ}$, and
$\left(N_{\rm R}^d\right)_{IJ}$ is 4.
In our case, they are automatically satisfied,
because the transformation matrices are given in the form as $\tilde{V} N$
where $\tilde{V}$ is an arbitrary unitary matrix and $N$ is a block-diagonal
matrix with $N_{I5} = 0$ and $N_{5J} = 0$.

Next, from the relations (\ref{M(U)-diag}) and (\ref{M(D)-diag}),
the following relations for the Yukawa couplings are obtained,
\begin{eqnarray}
&~& y_{IJ}^{(U)} = \lambda^7 \left({N_{\rm L}^u}^{\dagger}\right)_{I1} 
\left(N_{\rm R}^u\right)_{1J}
+ \lambda^4 \left({N_{\rm L}^u}^{\dagger}\right)_{I2} \left(N_{\rm R}^u\right)_{2J}
+ \lambda^0 \left({N_{\rm L}^u}^{\dagger}\right)_{I3} \left(N_{\rm R}^u\right)_{3J}
\nonumber \\
&~& ~~~~~~~~~~~~~ + \lambda^{-n_1} \left({N_{\rm L}^u}^{\dagger}\right)_{I4}
 \left(N_{\rm R}^u\right)_{5J}
+ \lambda^{-n_2} \left({N_{\rm L}^u}^{\dagger}\right)_{I5} \left(N_{\rm R}^u\right)_{4J},
\label{yuIJ}\\
&~& y^{(u_{\rm m})} 
= \lambda^7 \left({N_{\rm L}^u}^{\dagger}\right)_{51} \left(N_{\rm R}^u\right)_{15}
+ \lambda^4 \left({N_{\rm L}^u}^{\dagger}\right)_{52} \left(N_{\rm R}^u\right)_{25}
+ \lambda^0 \left({N_{\rm L}^u}^{\dagger}\right)_{53} \left(N_{\rm R}^u\right)_{35}
\nonumber \\
&~& ~~~~~~~~~~~~~ + \lambda^{-n_1} \left({N_{\rm L}^u}^{\dagger}\right)_{54} 
\left(N_{\rm R}^u\right)_{55}
+ \lambda^{-n_2} \left({N_{\rm L}^u}^{\dagger}\right)_{55} \left(N_{\rm R}^u\right)_{45},
\label{yu55}\\
&~& y_{IJ}^{(D)} = \lambda^7 \left({N_{\rm L}^d}^{\dagger}\right)_{I1} 
\left(N_{\rm R}^d\right)_{1J}
+ \lambda^5 \left({N_{\rm L}^d}^{\dagger}\right)_{I2} \left(N_{\rm R}^d\right)_{2J}
+ \lambda^3 \left({N_{\rm L}^d}^{\dagger}\right)_{I3} \left(N_{\rm R}^d\right)_{3J}
\nonumber \\
&~& ~~~~~~~~~~~~~ + \lambda^{-n_3} \left({N_{\rm L}^d}^{\dagger}\right)_{I4} 
\left(N_{\rm R}^d\right)_{5J}
+ \lambda^{-n_4} \left({N_{\rm L}^d}^{\dagger}\right)_{I5} \left(N_{\rm R}^d\right)_{4J},
\label{ydIJ}\\
&~& y^{(d_{\rm m})} 
= \lambda^7 \left({N_{\rm L}^d}^{\dagger}\right)_{51} \left(N_{\rm R}^d\right)_{15}
+ \lambda^5 \left({N_{\rm L}^d}^{\dagger}\right)_{52} \left(N_{\rm R}^d\right)_{25}
+ \lambda^3 \left({N_{\rm L}^d}^{\dagger}\right)_{53} \left(N_{\rm R}^d\right)_{35}
\nonumber \\
&~& ~~~~~~~~~~~~~ + \lambda^{-n_3} \left({N_{\rm L}^d}^{\dagger}\right)_{54} 
\left(N_{\rm R}^d\right)_{55}
+ \lambda^{-n_4} \left({N_{\rm L}^d}^{\dagger}\right)_{55} \left(N_{\rm R}^d\right)_{45}.
\label{yd55}
\end{eqnarray}
From Eq. (\ref{yuIJ}),
we find that $y_{IJ}^{(U)} = O(1)$
can be realized if a large mixing occurs between $u'_{{\rm L}3}$ ($u'_{{\rm R}3}$)
and other $u'_{{\rm L}}$s ($u'_{{\rm R}}$s), i.e.,
$\left({N_{\rm L}^u}^{\dagger}\right)_{I3} = O(1)$
and $\left(N_{\rm R}^u\right)_{3J} = O(1)$,
and even if other contributions are not sizable.
In contrast, for the down-type Yukawa couplings,
we have the following no-go theorem.\\
$[$Theorem$]$ {\it If the magnitude of kinetic coefficients is at most $O(1)$,
the magnitude of elements of down-type Yukawa coupling matrices cannot go beyond
$\displaystyle{O\left(10^{-2}\right)}$
without sizable contributions from extra fermions,
in the extension of the SM with vector-like fermions}.

In the following, we study a case with sizable contributions from extra quarks.
By imposing on
the conditions $y_{IJ}^{(D)} = O(1)$ and $y^{(d_{\rm m})}=O(1)$,
we obtain the relations:
\begin{eqnarray}
&~& \left({N_{\rm L}^d}^{\dagger}\right)_{I4} \left(N_{\rm R}^d\right)_{5J} 
= O\left(\lambda^{n_3}\right)~~~ {\rm and/or}~~~
\left({N_{\rm L}^d}^{\dagger}\right)_{I5} \left(N_{\rm R}^d\right)_{4J} 
= O\left(\lambda^{n_4}\right),
\label{NdI4}\\
&~& \left({N_{\rm L}^d}^{\dagger}\right)_{54} \left(N_{\rm R}^d\right)_{55} 
= O\left(\lambda^{n_3}\right)~~~ {\rm and/or}~~~
\left({N_{\rm L}^d}^{\dagger}\right)_{55} \left(N_{\rm R}^d\right)_{45} 
= O\left(\lambda^{n_4}\right).
\label{Nd54}
\end{eqnarray}
Hereafter, we consider the case with ``and'' in Eqs. (\ref{NdI4}) and (\ref{Nd54}).
Then, from $\left({N_{\rm L}^d}^{\dagger}\right)_{44} = O(1)$, 
$\left(N_{\rm R}^d\right)_{44} = O(1)$, $\left(N_{\rm R}^d\right)_{55} = O(1)$,
and $\left({N_{\rm L}^d}^{\dagger}\right)_{55} = O(1)$,
we obtain $\displaystyle{\left(N_{\rm R}^d\right)_{5J} = O\left(\lambda^{n_3}\right)}$, 
$\displaystyle{\left({N_{\rm L}^d}^{\dagger}\right)_{I5} = O\left(\lambda^{n_4}\right)}$,
$\displaystyle{\left({N_{\rm L}^d}^{\dagger}\right)_{54} = O\left(\lambda^{n_3}\right)}$,
and $\displaystyle{\left(N_{\rm R}^d\right)_{45} = O\left(\lambda^{n_4}\right)}$,
respectively.
Then, the transformation matrices take the form:
\begin{eqnarray}
{N_{\rm L}^d}^{\dagger}
= \left(
\begin{array}{ccccc}
1 & \star & \star & 1 & \lambda^{n_4} \\
\star & 1 & \star & 1 & \lambda^{n_4} \\
\star & \star & 1 & 1 & \lambda^{n_4} \\
\star & \star & \star & 1 & \lambda^{n_4} \\
\star & \star & \star & \lambda^{n_3} & 1 
\end{array} 
\right),~~~
N_{\rm R}^d
= \left(
\begin{array}{ccccc}
1 & \star & \star & \star & \star \\
\star & 1 & \star & \star & \star \\
\star & \star & 1 & \star & \star \\
1 & 1 & 1 & 1 & \lambda^{n_4} \\
\lambda^{n_3} & \lambda^{n_3} & \lambda^{n_3} & \lambda^{n_3} & 1 
\end{array} 
\right),
\label{NLd}
\end{eqnarray}
where 1 means $O(1)$ and $\star$ stands for an unspecified one.
From $K_{I5}^{(D_{\rm L})} 
= \left({N_{\rm L}^d}^{\dagger}\right)_{IA}
\left(N_{\rm L}^d\right)_{A5} = 0$
and $K_{5J}^{(D_{\rm L})} = \left({N_{\rm L}^d}^{\dagger}\right)_{5A}
\left(N_{\rm L}^d\right)_{AJ} = 0$,
we need $n_3 = n_4$ and 
$\displaystyle{\left({N_{\rm L}^d}^{\dagger}\right)_{5j} 
= O\left(\lambda^{n_3}\right)}$ ($j=1,2,3$).
From $K_{I5}^{(D_{\rm R})} 
= \left({N_{\rm R}^d}^{\dagger}\right)_{IA}
\left(N_{\rm R}^d\right)_{A5} = 0$
and $K_{5J}^{(D_{\rm R})} = \left({N_{\rm R}^d}^{\dagger}\right)_{5A}
\left(N_{\rm R}^d\right)_{AJ} = 0$,
we need $n_3 = n_4$ and 
$\displaystyle{\left({N_{\rm R}^d}\right)_{i5} = O\left(\lambda^{n_4}\right)}$ ($i=1,2,3$).
Then, we obtain the transformation matrices such as
\begin{eqnarray}
{N_{\rm L}^d},~~
N_{\rm R}^d
= \left(
\begin{array}{ccccc}
1 & \star & \star & \star & \lambda^{n_3} \\
\star & 1 & \star & \star & \lambda^{n_3} \\
\star & \star & 1 & \star & \lambda^{n_3} \\
1 & 1 & 1 & 1 & \lambda^{n_3} \\
\lambda^{n_3} & \lambda^{n_3} & \lambda^{n_3} & \lambda^{n_3} & 1 
\end{array} 
\right).
\label{NLd2}
\end{eqnarray}
In this way, we find that {\it an existence of a large mixing 
between $d'_{{\rm L}4}$ $(d'_{{\rm R}4})$
and other $d'_{{\rm L}}$s $(d'_{{\rm R}})$s, i.e.,
$\left({N_{\rm L}^d}^{\dagger}\right)_{I4} = O(1)$
and $\left(N_{\rm R}^d\right)_{4J} = O(1)$,
is necessary to generate $y^{(D)}_{IJ} = O(1)$.} 

Next, we examine the case that a magnitude of both 4th and 5th terms 
in $y_{IJ}^{(U)}$ and $y^{(u_{\rm m})}$ can be $O(1)$.
In this case, from $\left({N_{\rm L}^u}^{\dagger}\right)_{44} = O(1)$, 
$\left(N_{\rm R}^u\right)_{44} = O(1)$, 
$\left(N_{\rm R}^u\right)_{55} = O(1)$, and
$\left({N_{\rm L}^u}^{\dagger}\right)_{55} = O(1)$,
we obtain $\displaystyle{\left(N_{\rm R}^u\right)_{5J} = O\left(\lambda^{n_1}\right)}$,
$\displaystyle{\left({N_{\rm L}^u}^{\dagger}\right)_{I5} = O\left(\lambda^{n_2}\right)}$,
$\displaystyle{\left({N_{\rm L}^u}^{\dagger}\right)_{54} = O\left(\lambda^{n_1}\right)}$,
and
$\displaystyle{\left(N_{\rm R}^u\right)_{45} = O\left(\lambda^{n_2}\right)}$,respectively.
From the conditions $K_{I5}^{(U_{\rm L})} 
= \left({N_{\rm L}^u}^{\dagger}\right)_{IA}
\left(N_{\rm L}^u\right)_{A5} = 0$,
$K_{5J}^{(U_{\rm L})} = \left({N_{\rm L}^u}^{\dagger}\right)_{5A}
\left(N_{\rm L}^u\right)_{AJ} = 0$,
$K_{I5}^{(U_{\rm R})} 
= \left({N_{\rm R}^u}^{\dagger}\right)_{IA}
\left(N_{\rm R}^u\right)_{A5} = 0$,
and
$K_{5J}^{(U_{\rm R})} = \left({N_{\rm R}^u}^{\dagger}\right)_{5A}
\left(N_{\rm R}^u\right)_{AJ} = 0$,
we need $n_1 = n_2$, 
$\displaystyle{\left({N_{\rm L}^u}^{\dagger}\right)_{5j} = O\left(\lambda^{n_1}\right)}$,
and
$\displaystyle{\left({N_{\rm R}^u}\right)_{i5} = O\left(\lambda^{n_2}\right)}$.
Then, we obtain the transformation matrices such as
\begin{eqnarray}
{N_{\rm L}^u},~~
N_{\rm R}^u
= \left(
\begin{array}{ccccc}
1 & \star & \star & \star & \lambda^{n_1} \\
\star & 1 & \star & \star & \lambda^{n_1} \\
\star & \star & 1 & \star & \lambda^{n_1} \\
1 & 1 & 1 & 1 & \lambda^{n_1} \\
\lambda^{n_1} & \lambda^{n_1} & \lambda^{n_1} & \lambda^{n_1} & 1 
\end{array} 
\right).
\label{NLu}
\end{eqnarray}

For large masses concerning extra quarks, we derive the relations:
\begin{eqnarray}
&~& M_{I5}^{(U)} 
= \left\{\lambda^7 \left({N_{\rm L}^u}^{\dagger}\right)_{I1} \left(N_{\rm R}^u\right)_{15}
+ \lambda^4 \left({N_{\rm L}^u}^{\dagger}\right)_{I2} \left(N_{\rm R}^u\right)_{25}
+ \lambda^0 \left({N_{\rm L}^u}^{\dagger}\right)_{I3} \left(N_{\rm R}^u\right)_{35}\right.
\nonumber \\
&~& ~~~~~~~~~~~~~~~ 
\left. + \lambda^{-n_1} \left({N_{\rm L}^u}^{\dagger}\right)_{I4} \left(N_{\rm R}^u\right)_{55}
+ \lambda^{-n_1} \left({N_{\rm L}^u}^{\dagger}\right)_{I5} \left(N_{\rm R}^u\right)_{45}\right\}
\frac{v}{\sqrt{2}},
\label{MUI5}\\
&~& M_{I5}^{(D)} 
= \left\{\lambda^7 \left({N_{\rm L}^d}^{\dagger}\right)_{I1} \left(N_{\rm R}^d\right)_{15}
+ \lambda^5 \left({N_{\rm L}^d}^{\dagger}\right)_{I2} \left(N_{\rm R}^d\right)_{25}
+ \lambda^3 \left({N_{\rm L}^d}^{\dagger}\right)_{I3} \left(N_{\rm R}^d\right)_{35}\right.
\nonumber \\
&~& ~~~~~~~~~~~~~~~ 
\left. + \lambda^{-n_3} \left({N_{\rm L}^d}^{\dagger}\right)_{I4} \left(N_{\rm R}^d\right)_{55}
+ \lambda^{-n_3} \left({N_{\rm L}^d}^{\dagger}\right)_{I5} \left(N_{\rm R}^d\right)_{45}\right\}
\frac{v}{\sqrt{2}}.
\label{MDI5}
\end{eqnarray}
From the fact that the 4th terms in the right hand side of Eqs. (\ref{MUI5}) and (\ref{MDI5})
dominate over others and 
the consequence of SU(2)$_{\rm L}$ symmetry, i. e.,
$M_{I5}^{(U)} = M_{I5}^{(D)} = m_I^{(q_{\rm m})}$,
we obtain $n_1 = n_3$,
and then the following relation holds
\begin{eqnarray}
m_1^{(U)}
\left({N_{\rm L}^u}^{\dagger}\right)_{I4} \left(N_{\rm R}^u\right)_{55}
= m_1^{(D)}\left({N_{\rm L}^d}^{\dagger}\right)_{I4} \left(N_{\rm R}^d\right)_{55},
\label{MU=MD}
\end{eqnarray}
up to $\displaystyle{O\left(\lambda^{n_1} v/\sqrt{2}\right)}$.

When taken together, 
we have arrived at the transformation matrices 
to realize the conjecture (\ref{y=O(1)}), such as
\begin{eqnarray}
{N_{\rm L}^u},~~ {N_{\rm L}^d},~~ N_{\rm R}^u,~~ N_{\rm R}^d
= \left(
\begin{array}{ccccc}
1 & \star & \star & \star & \lambda^n \\
\star & 1 & \star & \star & \lambda^n \\
\star & \star & 1 & \star & \lambda^n \\
1 & 1 & 1 & 1 & \lambda^n \\
\lambda^n & \lambda^n & \lambda^n & \lambda^n & 1 
\end{array} 
\right),
\label{Ns}
\end{eqnarray}
under the parametrization (\ref{M-diag-rep}) with $n = n_1 = n_2 = n_3 = n_4$.
Inserting Eq. (\ref{Ns}) into Eq. (\ref{K(UL)}), 
we see that the magnitude of $k_{IJ}^{(q)}$, $k_{IJ}^{(u)}$, $k_{IJ}^{(d)}$,
$k^{(q_{\rm m})}$, $k^{(u_{\rm m})}$, and $k^{(d_{\rm m})}$
can become $O(1)$.

Finally, we study whether the Yukawa couplings can be unified or not.
As in the SM, we have the following no-go theorem.\\
$[$Theorem$]$ {\it Under the kinetic unification such as 
$k_{IJ}^{(q)} = k_{IJ}^{(u)} = k_{IJ}^{(d)}$ and
$k^{(q_{\rm m})} = k^{(u_{\rm m})} = k^{(d_{\rm m})}$
and the mass unification such as $m_J^{(u_{\rm m})} = m_J^{(d_{\rm m})}$,
the exact Yukawa coupling unification such as
$y^{(U)}_{IJ} = y^{(D)}_{IJ}$ does not occur
in the extension of the SM, described by ${\mathscr{L}'}_{\rm SM+\alpha}^{\rm quark}$}.

Here, we consider the case that the kinetic unification
and/or the mass unification are spoiled.
In this case, from Eqs. (\ref{yuIJ}) and (\ref{ydIJ}),
we find that $y^{(U)}_{IJ} = y^{(D)}_{IJ}$ is realized
with the relation:
\begin{eqnarray}
&~& m_u \left({N_{\rm L}^u}^{\dagger}\right)_{I1} \left(N_{\rm R}^u\right)_{1J}
+ m_c \left({N_{\rm L}^u}^{\dagger}\right)_{I2} \left(N_{\rm R}^u\right)_{2J}
+ m_t \left({N_{\rm L}^u}^{\dagger}\right)_{I3} \left(N_{\rm R}^u\right)_{3J}
\nonumber \\
&~& ~~~~~~~ + m_1^{(U)} \left({N_{\rm L}^u}^{\dagger}\right)_{I4} \left(N_{\rm R}^u\right)_{5J}
+ m_2^{(U)} \left({N_{\rm L}^u}^{\dagger}\right)_{I5} \left(N_{\rm R}^u\right)_{4J}
\nonumber \\
&~&   = m_d \left({N_{\rm L}^d}^{\dagger}\right)_{I1} \left(N_{\rm R}^d\right)_{1J}
+ m_s \left({N_{\rm L}^d}^{\dagger}\right)_{I2} \left(N_{\rm R}^d\right)_{2J}
+ m_b \left({N_{\rm L}^d}^{\dagger}\right)_{I3} \left(N_{\rm R}^d\right)_{3J}
\nonumber \\
&~& ~~~~~~~ + m_1^{(D)} \left({N_{\rm L}^d}^{\dagger}\right)_{I4} \left(N_{\rm R}^d\right)_{5J}
+ m_2^{(D)} \left({N_{\rm L}^d}^{\dagger}\right)_{I5} \left(N_{\rm R}^d\right)_{4J},
\label{yU=yD}
\end{eqnarray}
where mass parameters are replaced with values at the unification scale.
Here, we use quark masses defined by 
$m_u = y_u v/\sqrt{2}$, $m_c = y_c v/\sqrt{2}$, $m_t = y_t v/\sqrt{2}$,
$m_d = y_d v/\sqrt{2}$, $m_s = y_s v/\sqrt{2}$, and $m_b = y_b v/\sqrt{2}$. 
In general, there appear breaking terms that
contribute to the Yukawa coupling matrices,
on the breakdown of a higher gauge symmetry.
Then, the relation (\ref{yU=yD}) is modified
and their effects should be considered in a model-building.

Our results are summarized as follows.
There is a possibility that
a magnitude of elements of Yukawa coupling matrices
is $O(1)$ and the Yukawa couplings are unified 
at some high-energy scale,
if transformation matrices take a particular form such as Eq. (\ref{Ns}). 

\subsection{Features}

We examine features on Yukawa coupling matrices 
and transformation matrices inferred from our conjectures.

First, our matrices have a feature
that a small mixing of quark flavors 
between the weak and the mass eigenstates 
can occur as a result of
a cancellation between a large mixing in $N_{\rm L}^u$ and $N_{\rm L}^d$.
In our setup, the flavor mixing is defined by the matrix:
\begin{eqnarray}
N_{\rm mix} = \left({N_{\rm L}^{u}}^{\dagger}\right)^{-1} {N_{\rm L}^{d}}^{\dagger},
\label{Nmix}
\end{eqnarray}
and $N_{\rm mix}$ contains $V_{\rm KM}$
as the sub-matrix $\displaystyle{\left(N_{\rm mix}\right)_{ij}}$ ($i, j = 1, 2, 3$).

Next, our matrices show a feature that small Yukawa couplings in the SM have a high-sensitivity
for a small variation of matrix elements,
and it suggests that  
the quark mass hierarchy occurs as a consequence of
a fine tuning among large parameters.
This feature lies the other end of that obtained by a `stability' principle~\cite{St,DR}.
Here, the stability principle means that
a tiny dimensionless parameter should not be sensitive to the change of
matrix elements including it.
In our model, the Yukawa coupling matrices are expressed by
\begin{eqnarray}
y_{IJ}^{(U)} = y_{u_i} \left(T_i^{(u)}\right)_{IJ}
+ \xi_a^{(U)} \left(T_a^{(U)}\right)_{IJ},~~
y_{IJ}^{(D)} = y_{d_i} \left(T_i^{(d)}\right)_{IJ}
+ \xi_a^{(D)} \left(T_a^{(D)}\right)_{IJ},
\label{yudIJ}
\end{eqnarray}
where $y_{u_i} = (y_u, y_c, y_t)$, $y_{d_i} = (y_d, y_s, y_b)$,
$\xi_a^{(U)} = \sqrt{2} m_a^{(U)}/v$, $\xi_a^{(D)} = \sqrt{2} m_a^{(D)}/v$ ($a = 1, 2$)
are dimensionless parameters,
and $T_i^{(u)}$, $T_a^{(U)}$, $T_i^{(d)}$, and $T_a^{(D)}$ are 
$4 \times 4$ complex matrices given by
\begin{eqnarray}
\hspace{-1.3cm}&~& \left(T_i^{(u)}\right)_{IJ} = \left({N_{\rm L}^u}^{\dagger}\right)_{Ii} 
\left(N_{\rm R}^u\right)_{iJ},~~
\left(T_1^{(U)}\right)_{IJ} = \left({N_{\rm L}^u}^{\dagger}\right)_{I4}
 \left(N_{\rm R}^u\right)_{5J},~~
\left(T_2^{(U)}\right)_{IJ} =
\left({N_{\rm L}^u}^{\dagger}\right)_{I5} \left(N_{\rm R}^u\right)_{4J},
\label{Tu}\\
\hspace{-1.3cm}&~& \left(T_i^{(d)}\right)_{IJ} = \left({N_{\rm L}^d}^{\dagger}\right)_{Ii} 
\left(N_{\rm R}^d\right)_{iJ},~~
\left(T_1^{(D)}\right)_{IJ} = \left({N_{\rm L}^d}^{\dagger}\right)_{I4}
 \left(N_{\rm R}^d\right)_{5J},~~
\left(T_2^{(D)}\right)_{IJ} =
\left({N_{\rm L}^d}^{\dagger}\right)_{I5} \left(N_{\rm R}^d\right)_{4J}.
\label{Td}
\end{eqnarray}
The conditions that $y_{u_i}$ are stable
under a change of the $(I, J)$ element are given by
\begin{eqnarray}
\left|\frac{\delta y_{u_i}}{y_{u_i}}\right| 
\lesssim \left|\frac{\delta y_{IJ}^{(U)}}{y_{IJ}^{(U)}}\right|,~~
\left|\frac{\delta y_{d_i}}{y_{d_i}}\right|
\lesssim \left|\frac{\delta y_{IJ}^{(D)}}{y_{IJ}^{(D)}}\right|,~~ 
({\rm no~~\!summation~~\!on}~~i,I,J)
\label{yudIJ-st}
\end{eqnarray}
for $\delta y_{IJ}^{(U)} \ll y_{IJ}^{(U)}$
and $\delta y_{IJ}^{(D)} \ll y_{IJ}^{(D)}$.
In other words, the sensitivity of $y_{u_i}$ in $y_{IJ}^{(U)}$ 
and $y_{d_i}$ in $y_{IJ}^{(D)}$ is defined by
\begin{eqnarray}
\left(\varDelta_{y_{u_i}}\right)_{IJ} \equiv
\frac{\left|{\delta y_{u_i}}/{y_{u_i}}\right|}
{\left|{\delta y_{IJ}^{(U)}}/{y_{IJ}^{(U)}}\right|},~~
\left(\varDelta_{y_{d_i}}\right)_{IJ} \equiv
\frac{\left|{\delta y_{d_i}}/{y_{d_i}}\right|}
{\left|{\delta y_{IJ}^{(D)}}/{y_{IJ}^{(D)}}\right|},~~ ({\rm no~~\!summation~~\!on}~~i,I,J)
\label{yu-sen}
\end{eqnarray}
respectively, and then $\left(\varDelta_{y_{u_i}}\right)_{IJ} \le O(1)$ and 
$\left(\varDelta_{y_{d_i}}\right)_{IJ} \le O(1)$ are required from the stability condition.
If $\left(\varDelta_{y_u}\right)_{IJ} \gg O(1)$, $y_u$ has a high-sensitivity under the change
of other parameters. 

Using the Yukawa coupling matrices (\ref{yudIJ}), we derive the relations:
\begin{eqnarray}
&~& \frac{\delta y_{IJ}^{(U)}}{y_{IJ}^{(U)}}
= \frac{\delta y_{u_i}}{y_{u_i}} \frac{y_{u_i}\left(T_i^{(u)}\right)_{IJ}}{y_{IJ}^{(U)}}
+ \frac{\delta \xi_a^{(U)}}{\xi_a^{(U)}} 
\frac{\xi_a^{(U)}\left(T_a^{(U)}\right)_{IJ}}{{y_{IJ}^{(U)}}},~~
\label{delta-yu}\\
&~& \frac{\delta y_{IJ}^{(D)}}{y_{IJ}^{(D)}}
= \frac{\delta y_{d_i}}{y_{d_i}} \frac{y_{d_i}\left(T_i^{(d)}\right)_{IJ}}{y_{IJ}^{(D)}}
+ \frac{\delta \xi_a^{(D)}}{\xi_a^{(D)}}
\frac{\xi_a^{(D)}\left(T_a^{(D)}\right)_{IJ}}{{y_{IJ}^{(D)}}},
\label{delta-yd}
\end{eqnarray}
where no summations on $I$ and $J$ are done.
Using the relations (\ref{yudIJ-st}), (\ref{delta-yu}), and (\ref{delta-yd}),
we obtain the conditions:
\begin{eqnarray}
\left|y_{IJ}^{(U)}\right| \lesssim \left|y_{u_i}\left(T_i^{(u)}\right)_{IJ}\right|,~~
\left|y_{IJ}^{(D)}\right| \lesssim \left|y_{d_i}\left(T_i^{(d)}\right)_{IJ}\right|,
~~ ({\rm no~~\!summation~~\!on}~~i).
\label{|y|}
\end{eqnarray}
To fulfill the conditions (\ref{|y|}) for $y_u$ and $y_d$, 
the following inequalities are needed,
\begin{eqnarray}
&~& \left(T_2^{(u)}\right)_{IJ} \le O\left(\lambda^3\right),~~
\left(T_3^{(u)}\right)_{IJ} \le O\left(\lambda^7\right),~~
\left(T_a^{(U)}\right)_{IJ} \le O\left(\lambda^{7+n}\right),
\label{Tu-cond}\\
&~& \left(T_2^{(d)}\right)_{IJ} \le O\left(\lambda^2\right),~~
\left(T_3^{(d)}\right)_{IJ} \le O\left(\lambda^4\right),~~
\left(T_a^{(D)}\right)_{IJ} \le O\left(\lambda^{7+n}\right),
\label{Td-cond}
\end{eqnarray}
for the $(I, J)$ element with $\left(T_1^{(u)}\right)_{IJ}= O(1)$
and $\left(T_1^{(d)}\right)_{IJ}= O(1)$.
In this way, we conjecture that
the Yukawa couplings associated with a large mixing 
hardly satisfy the stability condition.

Finally, we give a speculation on the Yukawa coupling unification.
After terms containing $m_u$, $m_c$, $m_d$, $m_s$, and $m_b$ are neglected,
the relation (\ref{yU=yD}) is written by
\begin{eqnarray}
m_t \left(T_3^{(u)}\right)_{IJ}
+ m_1^{(U)} \left(T_1^{(U)}\right)_{IJ}
+ m_2^{(U)} \left(T_2^{(U)}\right)_{IJ}
=  m_1^{(D)} \left(T_1^{(D)}\right)_{IJ}
+ m_2^{(D)} \left(T_2^{(D)}\right)_{IJ}.
\label{yU=yD2}
\end{eqnarray}
For simplicity, we consider a case with $\left(T_1^{(U)}\right)_{IJ}=0$,
$\left(T_2^{(U)}\right)_{IJ} = 0$,
and $\left(T_2^{(D)}\right)_{IJ} = 0$.
In this case, the relation reduces to
\begin{eqnarray}
m_t \left({N_{\rm L}^u}^{\dagger}\right)_{I3} \left(N_{\rm R}^u\right)_{3J}
=  m_1^{(D)} \left({N_{\rm L}^d}^{\dagger}\right)_{I4} \left(N_{\rm R}^d\right)_{5J},
\label{yU=yD-1}
\end{eqnarray}
and this is realized by transformation matrices which satisfy
\begin{eqnarray}
\left({N_{\rm L}^u}^{\dagger}\right)_{I3} 
=\left({N_{\rm L}^d}^{\dagger}\right)_{I4},~~
m_t  \left(N_{\rm R}^u\right)_{3J}
=  m_1^{(D)} \left(N_{\rm R}^d\right)_{5J}.
\label{yU=yD-case1-sol}
\end{eqnarray}
They show that $d'_{{\rm L}4}$ and $\left(d'_{\rm L(m)}\right)^{\rm c}$
in the down-type quark sector
play a role of $u'_{{\rm L}3}$ and $u'_{{\rm R}3}$
in the up-type one, respectively.

In a case with $\left(T_1^{(U)}\right)_{IJ}=0$,
$\left(T_2^{(U)}\right)_{IJ} = 0$,
and $\left(T_1^{(D)}\right)_{IJ} = 0$,
the relation reduces to
\begin{eqnarray}
m_t \left({N_{\rm L}^u}^{\dagger}\right)_{I3} \left(N_{\rm R}^u\right)_{3J}
=  m_2^{(D)} \left({N_{\rm L}^d}^{\dagger}\right)_{I5} \left(N_{\rm R}^d\right)_{4J}.
\label{yU=yD-2}
\end{eqnarray}
and this is realized by transformation matrices which satisfy
\begin{eqnarray}
m_t  \left({N_{\rm L}^u}^{\dagger}\right)_{I3} 
= m_2^{(D)} \left({N_{\rm L}^d}^{\dagger}\right)_{I5},~~
\left(N_{\rm R}^u\right)_{3J}
= \left(N_{\rm R}^d\right)_{4J}.
\label{yU=yD-2-sol}
\end{eqnarray}
They show that $\left(d'_{\rm R(m)}\right)^{\rm c}$ and $d'_{{\rm R}4}$
in the down-type quark sector
play a role of $u'_{{\rm L}3}$ and $u'_{{\rm R}3}$
in the up-type one, respectively.
 
\section{Conclusions and discussions}

We have proposed a bottom-up approach that a structure of a high-energy physics
is explored by accumulating existence proofs and/or no-go theorems 
in the SM or its extension,
and studied an origin of fermion mass hierarchy 
based on an extension of the SM
with vector-like fermions, as an illustration.
It is shown that
the magnitude of elements of Yukawa coupling matrices
can be sizable of $O(1)$ and the Yukawa coupling unification 
can be realized at a high-energy scale, if vector-like fermions mix with
three families.
Through study on the extension with partial multiplets
of vector-like fermions (see Appendix B),
we have a no-go theorem such that
{\it the magnitude of elements of down-type Yukawa coupling matrices cannot go beyond
$\displaystyle{O\left(10^{-2}\right)}$,
in the extension of the SM
with vector-like up-type quarks alone, 
if the magnitude of kinetic coefficients is at most $O(1)$}.

Our results are obtained, independent of a theory beyond our setup,
and hence they could hold in various models.
In fact, large masses of extra fermions are free parameters,
and it is difficult to determine their magnitude 
from theoretical considerations alone.
Vector-like fermions including down-type ones can exist at a terascale
as remnants of unification and supersymmetry~\cite{YK3}. 
Conversely, if vector-like fermions are discovered 
and their masses are precisely measured, 
they would provide useful information about
our setup and take hints for HETs.

We explain preceding works on the fermion masses
based on models with vector-like fermions.
Vector-like fermions are used to generate small quark masses 
through a see-saw type mechanism~\cite{B,CM,DW,R}
and radiative corrections~\cite{BKM}.
The flavor structure has been studied in unified theories
with vector-like fermions~\cite{B2,BB,BB2,M,MYE6,BBT,DFM,BC}.
It would be meaningful to reexamine various models
with our conclusions in mind.
For a realistic model-building based on a grand unification,
we must consider leptons as well as quarks.
Our method is also applicable to the lepton sector
and models with several Higgs doublets.
In the case that vector-like fermions with heavy masses of the
unified scale are utilized, it would be convenient to treat unified theories
in a broken phase instead of BSM.

In our case with a large mixing in transformation matrices, 
Yukawa coupling matrices contain tiny parameters
such as $y_u$ and $y_d$ possessing a high-sensitivity
for a small variation of matrix elements, 
and hence the quark mass hierarchy can occur as a consequence of
accidental cancellations among large parameters.
Because this feature is different from that obtained by the stability principle,
it seems to be unnatural.
However, if transformation matrices with a large mixing
can be connected with ones with a small mixing by global symmetries 
in fermion kinetic terms in the similar manner as the SM,
unnatural features might be artificial.

Finally, we point out a limitation of our approach
and problems left behind.
From the bottom up, it is possible to show a possible existence
(if not an existence proof) and to offer some suggestions for HETs,
but it is difficult to specify the structure of HETs from our findings alone.
It would be a next task
to answer the following questions by exploiting to various methods.
Are transformation matrices with a large mixing realistic? 
Is there any circumstantial evidence to support them?
Or is there a model, theory or mechanism to realize them?

At present, a theory beyond the SM has not been yet known,
and hence it is worth pursuing every possibility
including a large mixing or a fine tuning.
Then, our approach would be useful as a complimentary one
to solve riddles in the SM.

\section*{Acknowledgments}
The author thanks Prof. T.~Yamashita for valuable discussions.
This work was supported in part by scientific grants 
from the Ministry of Education, Culture,
Sports, Science and Technology under Grant No.~17K05413.

\appendix

\section{Backgrounds of conjecture}

We explain the backgrounds of our first conjecture.
We assume that a fundamental theory originally contains no parameters,
and parameters $p_k$ emerge as the VEVs of 
some scalar fields such as moduli fields $\phi_k^{({\rm M})}$, 
i.e., $\displaystyle{p_k = \left\langle \phi_k^{({\rm M})} \right\rangle/M}$.
Here, $M$ is a fundamental scale such as a string scale or the Planck scale.
The magnitude of $p_k$ should be $O(1)$ 
if that of $\displaystyle{\left\langle \phi_k^{({\rm M})} \right\rangle}$ is $O(M)$.
Its high-energy effective theory (HET) can be described by the Lagrangian density:
\begin{eqnarray}
\mathscr{L}_{\rm HET} 
= \sum_i \frac{c_i(p_k)}{M^{n_i}} O_i(\psi, \phi, \varphi),
\label{L-HET}
\end{eqnarray}
where $c_i(p_k)$ are dimensionless coupling constants
and $O_i(\psi, \phi, \varphi)$ are operators with mass-dimension $(4+n_i)$.
Here, the magnitude of $c_i(p_k)$ can be $O(1)$
in the absence of a fine tuning.
Note that this feature is only a conjecture,
and it does not necessarily hold in the presence of tiny VEVs of moduli fields and/or
a fine tuning.
The HET turns out to be a theory beyond the SM (BSM), 
after some symmetries are broken down by the VEVs of
some scalar fields $\varphi_l$ at some high-energy scale $M_{\rm U}$
(less than $M$).
Then, the BSM can be described by the Lagrangian density:
\begin{eqnarray}
\mathscr{L}_{\rm BSM} 
= \sum_i \frac{c^{\rm BSM}_i(p_k, \varepsilon_l)}{M_{\rm U}^{n_i}} O^{\rm BSM}_i(\psi, \phi),
\label{L-BSM}
\end{eqnarray}
where $c^{\rm BSM}_i(p_k, \varepsilon_l)$ are dimensionless coupling constants,
$\varepsilon_l$ are dimensionless parameters defined by
$\displaystyle{\varepsilon_l \equiv \left\langle \varphi_l \right\rangle/M}$,
and $O^{\rm BSM}_i(\psi, \phi)$ are operators with mass-dimension $(4+n_i)$.
Note that there appear non-renormalizable terms suppressed by the power of $M_{\rm U}$,
after integrating out heavy fields with masses of $O(M_{\rm U})$.
The magnitude of $\varepsilon_l$ can be less than $O(1)$, 
when $\displaystyle{\left\langle \varphi_l \right\rangle = O(M_{\rm U}) < M}$.
Then, if dominant parts in $c^{\rm BSM}_i(p_k, \varepsilon_l)$
do not contain $\varepsilon_l$, 
the magnitude of $c^{\rm BSM}_i(p_k, \varepsilon_l)$ can be $O(1)$
in the absence of a fine tuning.
On the other hand, if dominant parts in $c^{\rm BSM}_i(p_k, \varepsilon_l)$ 
contain $\varepsilon_l$, 
the magnitude of $c^{\rm BSM}_i(p_k, \varepsilon_l)$ can be less than $O(1)$.

In HET, the flavor bases of matter fields $\psi'_I$ 
are defined by the canonical kinetic terms,
and kinetic terms and would-be kinetic terms are written by
\begin{eqnarray}
\sum_I \overline{\psi}'_I i \SlashD \psi'_I 
+ \sum_{I, J} K_{IJ}\left(p_k, \varphi_l\right) \overline{\psi}'_I i \SlashD \psi'_J.
\label{kin}
\end{eqnarray}
Then, the kinetic coefficients $k_{IJ}(p_k, \varepsilon_l)$ in BSM take the form:
\begin{eqnarray}
k_{IJ} = \delta_{IJ} + K_{IJ}\left(p_k,\left\langle \varphi_l \right\rangle\right) 
= \delta_{IJ} + c_{IJ}(p_k, \varepsilon_l).
\label{KIJ}
\end{eqnarray}
From Eq.~(\ref{KIJ}), we estimate that $k_{IJ} = O(1)$ for $I = J$
and $k_{IJ} \le O(1)$ for $I \ne J$, 
because off-diagonal components, in general, contain $\varepsilon_l (< O(1))$.
Here, we add an equality for $k_{IJ}$ ($I \ne J$)
by considering unknown non-perturbative contributions.
In contrast, the magnitude of Yukawa coupling matrices $y_{IJ}$ can be $O(1)$
if they originate from operators with mass-dimension 4 
or dominant parts of $y_{IJ}$ do not contain $\varepsilon_l$.

In an ordinary case, most studies have been carried out
based on a hierarchical type of Yukawa coupling matrices including a small parameter
or ansatzes called texture zeros,
based on canonical kinetic terms.
In this setup, all parameter regions can be covered 
by the re-definition of field variables, without loss of generality,
and fermion mass hierarchies can be easily generated
without a severe fine tuning.
Then, such attempts are plausible
to know the structure of HET in a broken phase,
described by canonical bases of fermions.

Let us advance one more step toward the understanding of the structure
of HET in an unbroken phase, e.g.,(\ref{L-HET}) and (\ref{kin}).
In this trial, it is suitable to treat a generic case (including
the case with $y_{IJ} = O(1)$ for all elements
or $y_{IJ} \ll O(1)$ for all elements) in BSM, based on non-canonical kinetic terms.
In other words, it is convenient to use non-canonical variables
in identification of the structure of HET, 
matching between HET and BSM (or the SM),
although the matching conditions are imposed on parameters and fields
in the broken phase of HET,
if HET contains (would-be) kinetic terms other than the canonical form.

Here, we consider the case with $y_{IJ} = O(1)$ for all elements.
In HET, $y_{IJ}$ are, in general, composed of two parts with a different magnitude:
\begin{eqnarray}
y_{IJ} = y^{(1)}_{IJ}(p_k) + y^{(\varepsilon)}_{IJ}(p_k, \varepsilon_l),
\label{yIJ}
\end{eqnarray}
where $y^{(1)}_{IJ}(p_k) = O(1)$ and $y^{(\varepsilon)}_{IJ}(p_k, \varepsilon_l) \ll O(1)$.
The difference of magnitude remains, 
after non-canonical kinetic terms are changed into canonical ones
due to the change of field variables.
We consider a case with $n$ families
and also denote the Yukawa coupling matrices for canonical kinetic terms as $y_{IJ}$.
In this case, $y_{IJ}$ is an $n \times n$ matrix.
When the rank of $y^{(1)}_{IJ}$ is $r$, a splitting of eigenvalues occurs,
i.e., $y_{IJ}$ has $r$ eigenvalues of $O(1)$ and $n-r$ ones of order 
much less than $O(1)$.
This can be a seed of a mass hierarchy.
A typical example is a democratic type matrix 
whose component has an identical value and whose rank is 1.
We need a mechanism to generate another hierarchy among small masses.

From the above observation, we hit on the idea
that fermion mass hierarchies might come from an excellent mechanism
with $y_{IJ} = O(1)$, and have a following speculation.
From the difference of kinetic coefficients,
the difference between the up-type quark Yukawa coupling matrix $y_{IJ}^{(U)}$
and the down-type one $y_{IJ}^{(D)}$ can come in.
If the rank of $y^{(1)}_{IJ}$ part in $y_{IJ}^{(U)}$ is $n-2$ 
and that of $y_{IJ}^{(D)}$ is $n-3$,
there can appear 2 up-type quarks and 3 down-type quarks much below the weak scale.
The $n-3$ sets of up-type and down-type quarks form vector-like heavy fermions
in company with their mirror ones,
and a up-type quark with the Yukawa coupling of $O(1)$ remains
as a chiral one and is identified as a top quark.
This is a basic idea behind our study.

\section{Cases with partial multiplets}

\subsection{Case with $q'_{{\rm L}4}$, $q'_{{\rm L(m)}}$, 
$u'_{{\rm R}4}$, and $u'_{{\rm R(m)}}$}

In the absence of $d'_{{\rm R}4}$ and $d'_{{\rm R(m)}}$, 
the quark kinetic coefficients are written by
\begin{eqnarray}
&~& K^{(U_{\rm L})} = 
\left(
\begin{array}{cc}
k_{IJ}^{(q)} & 0 \\
0 & k^{(u_{\rm m})}
\end{array} 
\right) = {N_{\rm L}^{u}}^{\dagger}N_{\rm L}^{u},~~
K^{(D_{\rm L})} = k_{IJ}^{(q)}
 = {N_{\rm L}^{d}}^{\dagger}N_{\rm L}^{d},~~
\nonumber \\
&~& K^{(U_{\rm R})} = 
\left(
\begin{array}{cc}
k_{IJ}^{(u)} & 0 \\
0 & k^{(q_{\rm m})}
\end{array} 
\right) = {N_{\rm R}^{u}}^{\dagger}N_{\rm R}^{u},~~
K^{(D_{\rm R})} = 
\left(
\begin{array}{cc}
k_{ij}^{(d)} & 0 \\
0 & k^{(q_{\rm m})}
\end{array} 
\right) = {N_{\rm R}^{d}}^{\dagger}N_{\rm R}^{d},
\label{K(UL)-case1}
\end{eqnarray}
where $N_{\rm L}^{u}$ and $N_{\rm R}^{u}$
are $5 \times 5$ complex matrices, and 
$N_{\rm L}^{d}$ and $N_{\rm R}^{d}$
are $4 \times 4$ complex matrices.

Relations of mass matrices are written by
\begin{eqnarray}
&~& 
\left(
\begin{array}{cc}
y_{IJ}^{(U)} \langle \phi^{0*} \rangle & m_I^{(q_{\rm m})} \\
m_J^{(u_{\rm m})} & y^{(u_{\rm m})} \langle \phi^{0} \rangle
\end{array} 
\right)
= {N_{\rm L}^{u}}^{\dagger} 
\left(
\begin{array}{ccccc}
m_u & 0 & 0 & 0 & 0 \\
0 & m_c & 0 & 0 & 0 \\
0 & 0 & m_t & 0 & 0 \\
0 & 0 & 0 & 0 & m_1^{(U)} \\
0 & 0 & 0 & m_2^{(U)} & 0
\end{array} 
\right) N_{\rm R}^{u},
\label{M(U)-case1}\\
&~& 
\left(
\begin{array}{cc}
y_{Ij}^{(D)} \langle \phi^{0} \rangle & m_I^{(q_{\rm m})} 
\end{array} 
\right)
= {N_{\rm L}^{d}}^{\dagger}
\left(
\begin{array}{cccc}
m_d & 0 & 0 & 0  \\
0 & m_s & 0 & 0  \\
0 & 0 & m_b & 0  \\
0 & 0 & 0  & m^{(D)} 
\end{array} 
\right) N_{\rm R}^{d},
\label{M(D)-case1}
\end{eqnarray}
where the magnitude of $m_1^{(U)}$, $m_2^{(U)}$, and $m^{(D)}$
is given by $\displaystyle{O\left(\lambda^{-n} v/\sqrt{2}\right)}$.
Then, we obtain the following transformation matrices
which realize $\displaystyle{y_{IJ}^{(U)}, y_{Ij}^{(D)}, y^{(u_{\rm m})} =O(1)}$,
\begin{eqnarray}
&~& {N_{\rm L}^u},~~ N_{\rm R}^u
= \left(
\begin{array}{ccccc}
1 & \star & \star & \star & \lambda^{n} \\
\star & 1 & \star & \star & \lambda^{n} \\
\star & \star & 1 & \star & \lambda^{n} \\
1 & 1 & 1 & 1 & \lambda^{n} \\
\lambda^n & \lambda^n & \lambda^n & \lambda^{n} & 1 
\end{array} 
\right),~~~
\label{Nu-case1}\\
&~& {N_{\rm L}^d}^{\dagger}
= \left(
\begin{array}{cccc}
1 & \star & \star & 1  \\
\star & 1 & \star & 1  \\
\star & \star & 1 & 1  \\
\star & \star & \star & 1 
\end{array} 
\right),~~~
N_{\rm R}^d
= \left(
\begin{array}{cccc}
1 & \star & \star & \lambda^n  \\
\star & 1 & \star & \lambda^n  \\
\star & \star & 1 & \lambda^n  \\
\lambda^n & \lambda^n & \lambda^n & 1 
\end{array} 
\right),
\label{Nd-case1}
\end{eqnarray}
using Eqs. (\ref{K(UL)-case1}), (\ref{M(U)-case1}), and (\ref{M(D)-case1}).

The relation $\displaystyle{y_{Ij}^{(U)} = y_{Ij}^{(D)}}$ is realized if the following
relation holds,
\begin{eqnarray}
&~& m_t \left({N_{\rm L}^u}^{\dagger}\right)_{I3} \left(N_{\rm R}^u\right)_{3j}
+ m_1^{(U)} \left({N_{\rm L}^u}^{\dagger}\right)_{I4} \left(N_{\rm R}^u\right)_{5j}
+ m_2^{(U)} \left({N_{\rm L}^u}^{\dagger}\right)_{I5} \left(N_{\rm R}^u\right)_{4j}
\nonumber \\
&~& = m^{(D)} \left({N_{\rm L}^d}^{\dagger}\right)_{I4} \left(N_{\rm R}^d\right)_{4j},
\label{yU=yD-case1}
\end{eqnarray}
where we neglect tiny contributions including $m_u$, $m_c$, $m_d$, $m_s$, and $m_b$.

\subsection{Case with $q'_{{\rm L}4}$, $q'_{{\rm L(m)}}$, 
$d'_{{\rm R}4}$, and $d'_{{\rm R(m)}}$}

In the absence of $u'_{{\rm R}4}$ and $u'_{{\rm R(m)}}$, 
the quark kinetic coefficients are written by
\begin{eqnarray}
&~& K^{(U_{\rm L})} = k_{IJ}^{(q)}
 = {N_{\rm L}^{u}}^{\dagger}N_{\rm L}^{u},~~
K^{(D_{\rm L})} = \left(
\begin{array}{cc}
k_{IJ}^{(q)} & 0 \\
0 & k^{(d_{\rm m})}
\end{array} 
\right)
 = {N_{\rm L}^{d}}^{\dagger}N_{\rm L}^{d},~~
\nonumber \\
&~& K^{(U_{\rm R})} = 
\left(
\begin{array}{cc}
k_{ij}^{(u)} & 0 \\
0 & k^{(q_{\rm m})}
\end{array} 
\right) = {N_{\rm R}^{u}}^{\dagger}N_{\rm R}^{u},~~
K^{(D_{\rm R})} = 
\left(
\begin{array}{cc}
k_{IJ}^{(d)} & 0 \\
0 & k^{(q_{\rm m})}
\end{array} 
\right) = {N_{\rm R}^{d}}^{\dagger}N_{\rm R}^{d},
\label{K(UL)-case2}
\end{eqnarray}
where $N_{\rm L}^{u}$ and $N_{\rm R}^{u}$
are $4 \times 4$ complex matrices, and 
$N_{\rm L}^{d}$ and $N_{\rm R}^{d}$
are $5 \times 5$ complex matrices.

Relations of mass matrices are written by
\begin{eqnarray}
&~& 
\left(
\begin{array}{cc}
y_{Ij}^{(U)} \langle \phi^{0*} \rangle & m_I^{(q_{\rm m})} 
\end{array} 
\right)
= {N_{\rm L}^{u}}^{\dagger} 
\left(
\begin{array}{cccc}
m_u & 0 & 0 & 0  \\
0 & m_c & 0 & 0  \\
0 & 0 & m_t & 0  \\
0 & 0 & 0  & m^{(U)} 
\end{array} 
\right) N_{\rm R}^{u},
\label{M(U)-case2}\\
&~& 
\left(
\begin{array}{cc}
y_{IJ}^{(D)} \langle \phi^{0} \rangle & m_I^{(q_{\rm m})} \\
m_J^{(d_{\rm m})} & y^{(d_{\rm m})} \langle \phi^{0*} \rangle
\end{array} 
\right)
= {N_{\rm L}^{d}}^{\dagger}
\left(
\begin{array}{ccccc}
m_d & 0 & 0 & 0 & 0 \\
0 & m_s & 0 & 0 & 0 \\
0 & 0 & m_b & 0 & 0 \\
0 & 0 & 0 & 0 & m_1^{(D)} \\
0 & 0 & 0 & m_2^{(D)} & 0
\end{array} 
\right) N_{\rm R}^{d},
\label{M(D)-case2}
\end{eqnarray}
where the magnitude of $m^{(U)}$, $m_1^{(D)}$, and $m_2^{(D)}$
is given by $\displaystyle{O\left(\lambda^{-n} v/\sqrt{2}\right)}$.
Then, we obtain the following transformation matrices
which realize $\displaystyle{y_{Ij}^{(U)}, y_{IJ}^{(D)}, y^{(d_{\rm m})}=O(1)}$,
\begin{eqnarray}
&~& {N_{\rm L}^u}^{\dagger} 
= \left(
\begin{array}{cccc}
1 & \star & \star & 1  \\
\star & 1 & \star & 1  \\
\star & \star & 1 & 1  \\
\star & \star & \star & 1 
\end{array} 
\right),~~~
N_{\rm R}^u
= \left(
\begin{array}{cccc}
1 & \star & \star & \lambda^n  \\
\star & 1 & \star & \lambda^n  \\
\star & \star & 1 & \lambda^n  \\
\lambda^n & \lambda^n & \lambda^n & 1 
\end{array} 
\right),
\label{Nu-case2}\\
&~& {N_{\rm L}^d},~~ N_{\rm R}^d
= \left(
\begin{array}{ccccc}
1 & \star & \star & \star & \lambda^{n} \\
\star & 1 & \star & \star & \lambda^{n} \\
\star & \star & 1 & \star & \lambda^{n} \\
1 & 1 & 1 & 1 & \lambda^{n} \\
\lambda^n & \lambda^n & \lambda^n & \lambda^{n} & 1 
\end{array} 
\right),
\label{Nd-case2}
\end{eqnarray}
using Eqs. (\ref{K(UL)-case2}), (\ref{M(U)-case2}), and (\ref{M(D)-case2}).

The relation $\displaystyle{y_{Ij}^{(U)} = y_{Ij}^{(D)}}$ is realized if the following
relation holds,
\begin{eqnarray}
&~& m_t \left({N_{\rm L}^u}^{\dagger}\right)_{I3} \left(N_{\rm R}^u\right)_{3j}
+ m^{(U)} \left({N_{\rm L}^u}^{\dagger}\right)_{I4} \left(N_{\rm R}^u\right)_{4j}
\nonumber \\
&~& = m_1^{(D)} \left({N_{\rm L}^d}^{\dagger}\right)_{I4} \left(N_{\rm R}^d\right)_{5j}
+ m_2^{(D)} \left({N_{\rm L}^d}^{\dagger}\right)_{I5} \left(N_{\rm R}^d\right)_{4j},
\label{yU=yD-case2}
\end{eqnarray}
where we neglect tiny contributions including $m_u$, $m_c$, $m_d$, $m_s$, and $m_b$.

\subsection{Case with $u'_{{\rm R}4}$, $u'_{{\rm R(m)}}$, 
$d'_{{\rm R}4}$, and $d'_{{\rm R(m)}}$}

In the absence of $q'_{{\rm L}4}$ and $q'_{{\rm L(m)}}$, 
the quark kinetic coefficients are written by
\begin{eqnarray}
&~& K^{(U_{\rm L})} = 
\left(
\begin{array}{cc}
k_{ij}^{(q)} & 0 \\
0 & k^{(u_{\rm m})}
\end{array} 
\right) = {N_{\rm L}^{u}}^{\dagger}N_{\rm L}^{u},~~
K^{(D_{\rm L})} =
\left(
\begin{array}{cc}
k_{ij}^{(q)} & 0 \\
0 & k^{(d_{\rm m})}
\end{array} 
\right)  
 = {N_{\rm L}^{d}}^{\dagger}N_{\rm L}^{d},~~
\nonumber \\
&~& K^{(U_{\rm R})} = k_{IJ}^{(u)}
 = {N_{\rm R}^{u}}^{\dagger}N_{\rm R}^{u},~~
K^{(D_{\rm R})} = k_{IJ}^{(d)}
 = {N_{\rm R}^{d}}^{\dagger}N_{\rm R}^{d},
\label{K(UL)-case3}
\end{eqnarray}
where $N_{\rm L}^{u}$, $N_{\rm L}^{d}$, $N_{\rm R}^{u}$, and $N_{\rm R}^{d}$
are $4 \times 4$ complex matrices.

Relations of mass matrices are written by
\begin{eqnarray}
&~& 
\left(
\begin{array}{c}
y_{iJ}^{(U)} \langle \phi^{0*} \rangle  \\
m_J^{(u_{\rm m})}
\end{array} 
\right)
= {N_{\rm L}^{u}}^{\dagger} 
\left(
\begin{array}{cccc}
m_u & 0 & 0 & 0  \\
0 & m_c & 0 & 0  \\
0 & 0 & m_t & 0  \\
0 & 0 & 0 & m^{(U)} 
\end{array} 
\right) N_{\rm R}^{u},
\label{M(U)-case3}\\
&~& 
\left(
\begin{array}{c}
y_{iJ}^{(D)} \langle \phi^{0} \rangle \\
m_J^{(d_{\rm m})} 
\end{array} 
\right)
= {N_{\rm L}^{d}}^{\dagger}
\left(
\begin{array}{cccc}
m_d & 0 & 0 & 0  \\
0 & m_s & 0 & 0  \\
0 & 0 & m_b & 0  \\
0 & 0 & 0  & m^{(D)} 
\end{array} 
\right) N_{\rm R}^{d},
\label{M(D)-case3}
\end{eqnarray}
where the magnitude of $m^{(U)}$ and $m^{(D)}$
is given by $\displaystyle{O\left(\lambda^{-n} v/\sqrt{2}\right)}$.
Then, we obtain the following transformation matrices
which realize $\displaystyle{y_{iJ}^{(U)}, y_{iJ}^{(D)} =O(1)}$,
\begin{eqnarray}
{N_{\rm L}^u}^{\dagger},~~ {N_{\rm L}^d}^{\dagger}
= \left(
\begin{array}{cccc}
1 & \star & \star & \lambda^n  \\
\star & 1 & \star & \lambda^n  \\
\star & \star & 1 & \lambda^n  \\
\lambda^n & \lambda^n & \lambda^n & 1 
\end{array} 
\right),~~~ 
N_{\rm R}^u,~~ N_{\rm R}^d
= \left(
\begin{array}{cccc}
1 & \star & \star & 1 \\
\star & 1 & \star & 1 \\
\star & \star & 1 & 1 \\
1 & 1 & 1 & 1 
\end{array} 
\right),
\label{Ns-case3}
\end{eqnarray}
using (\ref{K(UL)-case3}), (\ref{M(U)-case3}), and (\ref{M(D)-case3}).

The relation $\displaystyle{y_{iJ}^{(U)} = y_{iJ}^{(D)}}$ is realized if the following
relation holds,
\begin{eqnarray}
m_t \left({N_{\rm L}^u}^{\dagger}\right)_{i3} \left(N_{\rm R}^u\right)_{3J}
+ m^{(U)} \left({N_{\rm L}^u}^{\dagger}\right)_{i4} \left(N_{\rm R}^u\right)_{4J}
= m^{(D)} \left({N_{\rm L}^d}^{\dagger}\right)_{i4} \left(N_{\rm R}^d\right)_{4J},
\label{yU=yD-case3}
\end{eqnarray}
where we neglect tiny contributions including $m_u$, $m_c$, $m_d$, $m_s$, and $m_b$.

\subsection{Case with $q'_{{\rm L}4}$ and $q'_{{\rm L(m)}}$}

In the absence of $u'_{{\rm R}4}$, $u'_{{\rm R(m)}}$, 
$d'_{{\rm R}4}$, and $d'_{{\rm R(m)}}$, the quark kinetic coefficients are written by
\begin{eqnarray}
&~& K^{(U_{\rm L})} = k_{IJ}^{(q)}
 = {N_{\rm L}^{u}}^{\dagger}N_{\rm L}^{u},~~
K^{(D_{\rm L})} = k_{IJ}^{(q)}
 = {N_{\rm L}^{d}}^{\dagger}N_{\rm L}^{d},~~
\nonumber \\
&~& K^{(U_{\rm R})} = 
\left(
\begin{array}{cc}
k_{ij}^{(u)} & 0 \\
0 & k^{(q_{\rm m})}
\end{array} 
\right) = {N_{\rm R}^{u}}^{\dagger}N_{\rm R}^{u},~~
K^{(D_{\rm R})} = 
\left(
\begin{array}{cc}
k_{ij}^{(d)} & 0 \\
0 & k^{(q_{\rm m})}
\end{array} 
\right) = {N_{\rm R}^{d}}^{\dagger}N_{\rm R}^{d},
\label{K(UL)-case4}
\end{eqnarray}
where $N_{\rm L}^{u}$, $N_{\rm L}^{d}$, $N_{\rm R}^{u}$, and $N_{\rm R}^{d}$
are $4 \times 4$ complex matrices.

Relations of mass matrices are written by
\begin{eqnarray}
&~& 
\left(
\begin{array}{cc}
y_{Ij}^{(U)} \langle \phi^{0*} \rangle & m_I^{(q_{\rm m})} 
\end{array} 
\right)
= {N_{\rm L}^{u}}^{\dagger} 
\left(
\begin{array}{cccc}
m_u & 0 & 0 & 0  \\
0 & m_c & 0 & 0  \\
0 & 0 & m_t & 0  \\
0 & 0 & 0 & m^{(U)} 
\end{array} 
\right) N_{\rm R}^{u},
\label{M(U)-case4}\\
&~& 
\left(
\begin{array}{cc}
y_{Ij}^{(D)} \langle \phi^{0} \rangle & m_I^{(q_{\rm m})} 
\end{array} 
\right)
= {N_{\rm L}^{d}}^{\dagger}
\left(
\begin{array}{cccc}
m_d & 0 & 0 & 0  \\
0 & m_s & 0 & 0  \\
0 & 0 & m_b & 0  \\
0 & 0 & 0  & m^{(D)} 
\end{array} 
\right) N_{\rm R}^{d},
\label{M(D)-case4}
\end{eqnarray}
where the magnitude of $m^{(U)}$ and $m^{(D)}$
is given by $\displaystyle{O\left(\lambda^{-n} v/\sqrt{2}\right)}$.
Then, we obtain the following transformation matrices
which realize $\displaystyle{y_{Ij}^{(U)}, y_{Ij}^{(D)} = O(1)}$,
\begin{eqnarray}
{N_{\rm L}^u}^{\dagger},~~ {N_{\rm L}^d}^{\dagger}
= \left(
\begin{array}{cccc}
1 & \star & \star & 1 \\
\star & 1 & \star & 1 \\
\star & \star & 1 & 1 \\
\star & \star & \star & 1 
\end{array} 
\right),~~~ 
N_{\rm R}^u,~~ N_{\rm R}^d
= \left(
\begin{array}{cccc}
1 & \star & \star & \lambda^n  \\
\star & 1 & \star & \lambda^n  \\
\star & \star & 1 & \lambda^n  \\
\lambda^n & \lambda^n & \lambda^n & 1 
\end{array} 
\right),
\label{Ns-case4}
\end{eqnarray}
using (\ref{K(UL)-case4}), (\ref{M(U)-case4}), and (\ref{M(D)-case4}).

The relation $\displaystyle{y_{Ij}^{(U)} = y_{Ij}^{(D)}}$ is realized if the following
relation holds,
\begin{eqnarray}
m_t \left({N_{\rm L}^u}^{\dagger}\right)_{I3} \left(N_{\rm R}^u\right)_{3j}
+ m^{(U)} \left({N_{\rm L}^u}^{\dagger}\right)_{I4} \left(N_{\rm R}^u\right)_{4j}
= m^{(D)} \left({N_{\rm L}^d}^{\dagger}\right)_{I4} \left(N_{\rm R}^d\right)_{4j},
\label{yU=yD-case4}
\end{eqnarray}
where we neglect tiny contributions including $m_u$, $m_c$, $m_d$, $m_s$, and $m_b$.

\subsection{Case with $u'_{{\rm R}4}$ and $u'_{{\rm R(m)}}$}

In the absence of $q'_{{\rm L}4}$, $q'_{{\rm L(m)}}$, 
$d'_{{\rm R}4}$, and $d'_{{\rm R(m)}}$, the quark kinetic coefficients are written by
\begin{eqnarray}
&~& K^{(U_{\rm L})} =\left(
\begin{array}{cc}
k_{ij}^{(q)} & 0 \\
0 & k^{(u_{\rm m})}
\end{array} 
\right) 
 = {N_{\rm L}^{u}}^{\dagger}N_{\rm L}^{u},~~
K^{(D_{\rm L})} = k_{ij}^{(q)}
 = {N_{\rm L}^{d}}^{\dagger}N_{\rm L}^{d},~~
\nonumber \\
&~& K^{(U_{\rm R})} = k_{IJ}^{(u)}
= {N_{\rm R}^{u}}^{\dagger}N_{\rm R}^{u},~~
K^{(D_{\rm R})} = k_{ij}^{(d)}
= {N_{\rm R}^{d}}^{\dagger}N_{\rm R}^{d},
\label{K(UL)-case5}
\end{eqnarray}
where $N_{\rm L}^{u}$ and $N_{\rm R}^{u}$
are $4 \times 4$ complex matrices, and 
$N_{\rm L}^{d}$ and $N_{\rm R}^{d}$
are $3 \times 3$ complex matrices.

Relations of mass matrices are written by
\begin{eqnarray}
&~& 
\left(
\begin{array}{c}
y_{iJ}^{(U)} \langle \phi^{0*} \rangle  \\
m_J^{(u_{\rm m})}
\end{array} 
\right)
= {N_{\rm L}^{u}}^{\dagger} 
\left(
\begin{array}{cccc}
m_u & 0 & 0 & 0  \\
0 & m_c & 0 & 0  \\
0 & 0 & m_t & 0  \\
0 & 0 & 0 & m^{(U)} 
\end{array} 
\right) N_{\rm R}^{u},
\label{M(U)-case5}\\
&~& 
y_{ij}^{(D)} \langle \phi^{0} \rangle 
= {N_{\rm L}^{d}}^{\dagger}
\left(
\begin{array}{ccc}
m_d & 0 & 0  \\
0 & m_s & 0  \\
0 & 0 & m_b  
\end{array} 
\right) N_{\rm R}^{d},
\label{M(D)-case5}
\end{eqnarray}
where the magnitude of $m^{(U)}$
is given by $\displaystyle{O\left(\lambda^{-n} v/\sqrt{2}\right)}$.
Then, there is no transformation matrices ${N_{\rm L}^d}^{\dagger}$
and $N_{\rm R}^d$ of $O(1)$ to realize $y_{ij}^{(D)} = O(1)$, as in the SM.

\subsection{Case with $d'_{{\rm R}4}$ and $d'_{{\rm R(m)}}$}

In the absence of $q'_{{\rm L}4}$, $q'_{{\rm L(m)}}$, 
$u'_{{\rm R}4}$, and $u'_{{\rm R(m)}}$, the quark kinetic coefficients are written by
\begin{eqnarray}
&~& K^{(U_{\rm L})} =k_{ij}^{(q)}
 = {N_{\rm L}^{u}}^{\dagger}N_{\rm L}^{u},~~
K^{(D_{\rm L})} = \left(
\begin{array}{cc}
k_{ij}^{(q)} & 0 \\
0 & k^{(d_{\rm m})}
\end{array} 
\right) 
 = {N_{\rm L}^{d}}^{\dagger}N_{\rm L}^{d},~~
\nonumber \\
&~& K^{(U_{\rm R})} = k_{ij}^{(u)}
= {N_{\rm R}^{u}}^{\dagger}N_{\rm R}^{u},~~
K^{(D_{\rm R})} = k_{IJ}^{(d)}
= {N_{\rm R}^{d}}^{\dagger}N_{\rm R}^{d},
\label{K(UL)-case6}
\end{eqnarray}
where $N_{\rm L}^{u}$ and $N_{\rm R}^{u}$
are $3 \times 3$ complex matrices, and 
$N_{\rm L}^{d}$ and $N_{\rm R}^{d}$
are $4 \times 4$ complex matrices.

Relations of mass matrices are written by
\begin{eqnarray}
&~& 
y_{ij}^{(U)} \langle \phi^{0*} \rangle
= {N_{\rm L}^{u}}^{\dagger} 
\left(
\begin{array}{ccc}
m_u & 0 & 0   \\
0 & m_c & 0   \\
0 & 0 & m_t   
\end{array} 
\right) N_{\rm R}^{u},
\label{M(U)-case6}\\
&~& 
\left(
\begin{array}{c}
y_{iJ}^{(D)} \langle \phi^{0} \rangle \\
m_J^{(u_{\rm m})}
\end{array} 
\right)
= {N_{\rm L}^{d}}^{\dagger}
\left(
\begin{array}{cccc}
m_d & 0 & 0 & 0 \\
0 & m_s & 0 & 0 \\
0 & 0 & m_b & 0 \\
0 & 0 & 0 & m^{(D)}
\end{array} 
\right) N_{\rm R}^{d},
\label{M(D)-case6}
\end{eqnarray}
where the magnitude of $m^{(D)}$
is given by $\displaystyle{O\left(\lambda^{-n} v/\sqrt{2}\right)}$.
Then, we obtain the following transformation matrices
which realize $\displaystyle{y_{ij}^{(U)}, y_{iJ}^{(D)} =O(1)}$,
\begin{eqnarray}
&~& {N_{\rm L}^u}^{\dagger}
= \left(
\begin{array}{ccc}
1 & \star & 1  \\
\star & 1 & 1  \\
\star & \star & 1   
\end{array} 
\right),~~~ N_{\rm R}^u
= \left(
\begin{array}{ccc}
1 & \star & \star  \\
\star & 1 & \star  \\
1 & 1 & 1   
\end{array} 
\right),
\label{Nu-case6}\\
&~& {N_{\rm L}^d}^{\dagger}
= \left(
\begin{array}{cccc}
1 & \star & \star & \lambda^{n} \\
\star & 1 & \star & \lambda^{n} \\
\star & \star & 1 & \lambda^{n} \\
\lambda^{n} & \lambda^{n} & \lambda^{n} & 1 
\end{array} 
\right),~~~ N_{\rm R}^d
= \left(
\begin{array}{cccc}
1 & \star & \star & \star \\
\star & 1 & \star & \star \\
\star & \star & 1 & \star \\
1 & 1 & 1 & 1 
\end{array} 
\right),
\label{Nd-case6}
\end{eqnarray}
using (\ref{K(UL)-case6}), (\ref{M(U)-case6}), and (\ref{M(D)-case6}).

The relation $\displaystyle{y_{ij}^{(U)} = y_{ij}^{(D)}}$ is realized if the following
relation holds,
\begin{eqnarray}
m_t \left({N_{\rm L}^u}^{\dagger}\right)_{i3} \left(N_{\rm R}^u\right)_{3j}
= m^{(D)} \left({N_{\rm L}^d}^{\dagger}\right)_{i4} \left(N_{\rm R}^d\right)_{4j},
\label{yU=yD-case6}
\end{eqnarray}
where we neglect tiny contributions including $m_u$, $m_c$, $m_d$, $m_s$, and $m_b$.
As a simple case, the above relation holds with 
\begin{eqnarray}
m_t \left({N_{\rm L}^u}^{\dagger}\right)_{i3} 
= m^{(D)} \left({N_{\rm L}^d}^{\dagger}\right)_{i4},~~
\left(N_{\rm R}^u\right)_{3j}
= \left(N_{\rm R}^d\right)_{4j}.
\label{yU=yD-sol}
\end{eqnarray}
In this case, $\left(d'_{\rm R(m)}\right)^{\rm c}$ and $d'_{{\rm R}4}$
in the down-type quark sector
play a role of $u'_{{\rm L}3}$ and $u'_{{\rm R}3}$
in the up-type one, respectively.

\end{document}